\documentclass[journal,twocolumn]{IEEEtran}
\usepackage[margin=1in]{geometry}
\usepackage{amsmath}
\usepackage{amsthm} 
\usepackage{amsfonts}
\usepackage{graphicx}
\graphicspath{{./images/}}

\usepackage{caption}
\usepackage{subcaption} 
\usepackage[monochrome]{xcolor}
\usepackage{algpseudocode}
\usepackage{algorithm}
\definecolor{kthblue}{rgb}{0.098, 0.329, 0.651}
\definecolor{newgreen}{RGB}{38, 115, 77}
\definecolor{newpurple}{RGB}{102, 0, 204}
\definecolor{newred}{RGB}{229, 0, 76}
\definecolor{newbrown}{RGB}{115, 77, 38}
\definecolor{neworange}{RGB}{255, 153, 0}
\usepackage{mathtools}
\usepackage[final]{changes}
\usepackage{url}

\title{\textcolor{newred}{Navigating chemical design spaces for metal-ion batteries via machine-learning-guided phase-field simulations}}

\author{\IEEEauthorblockN{Hamed Taghavian\IEEEauthorrefmark{2},
Viktor Vanoppen\IEEEauthorrefmark{2}, Erik Berg\IEEEauthorrefmark{2},
Peter Broqvist\IEEEauthorrefmark{2}\IEEEauthorrefmark{1} and Jens Sjölund\IEEEauthorrefmark{2}\IEEEauthorrefmark{1}}\\
\IEEEauthorblockA{\IEEEauthorrefmark{2}Uppsala University, Uppsala, Sweden
\\
Email: jens.sjolund@it.uu.se}
\\
Published version: https://rdcu.be/exHzo}

\begin{document}
\maketitle

\begin{abstract}
\textcolor{newred}{Metal anodes provide the highest energy density in batteries. However, they still suffer from electrode/electrolyte interface side reactions and dendrite growth, especially under fast-charging conditions.
In this paper, we consider a phase-field model of electrodeposition in metal-anode batteries and provide a scalable, versatile framework for optimizing its chemical parameters. Our approach is based on Bayesian optimization and explores the parameter space with a high sample efficiency and a low computation complexity. We use this framework to find the optimal cell for suppressing dendrite growth and accelerating charging speed under constant voltage. We identify interfacial mobility as a key parameter, which should be maximized to inhibit dendrites without compromising the charging speed. The results are verified using extended simulations of dendrite evolution in charging half cells with lithium-metal anodes.}
\end{abstract}



\section{INTRODUCTION}\label{sec:intro}
Metal plating is the holy grail of rechargeable batteries, providing cells with the highest possible energy density. However, dendrite formation hinders the widespread use of these batteries, reduces their lifespan, and poses safety risks. Controlling dendrite growth by optimizing the batteries' chemical parameters is essential to address these challenges. Performing this optimization through real-world experiments is not only exceedingly time-consuming but also restricted by the challenges of precisely controlling the chemical parameters. Meanwhile, analytical expressions are too simplistic to capture the complex multi-scale nature of dendrite formation. Simulations, in particular ones based on phase-field models, strike a middle ground: offering a tractable yet physically realistic approach to a fundamental understanding of dendrite evolution in battery cells. These models have been proven useful in understanding the influence of several different parameters of batteries on dendrite growth. {For example, phase-field models have shown that dendrite formation is inhibited when the applied voltage and exchange current density are low~\cite{zinc3}, the interface 
thickness is large~\cite{thickness}, the separator has a small pore size~\cite{separator_effect} and there is high external pressure on the anode~\cite{pressure1}. Phase-field models have also been used to describe the more nuanced effects of anisotropic strength and temperature on dendrite growth~\cite{Sodium,mueffect}.} 

Understanding the relation between different chemical and physical parameters of a battery and dendrite growth in phase-field models often relies on a complete simulation of dendrite evolution, by solving a system of nonlinear partial differential equations using the finite element method. {While this approach is useful for investigating the effects of a single parameter on dendrite growth, it does not scale well with the number of influencing parameters. A significant amount of time and computational resources are needed to investigate the simultaneous effects of several parameters on dendrite growth. As there is a large number of parameters in phase-field models with an unknown effect on dendrites, a systematic approach is required to find the optimal set of parameters that inhibits dendrite growth.

\textcolor{newred}{Furthermore, minimizing only dendrite growth is a restrictive design objective that can lead to conservative solutions, and in extreme cases, to trivial solutions that stop charging completely. The reason is that dendrite suppression and fast charging often have a conflicting nature. The growth of dendrites in a cell increases the surface area of the electrode, which offers more available spots for deposition, potentially enhancing the charging speed. This inverse relation between dendrite inhibition and fast charging manifests itself as health-related constraints that prevent large charging currents in fast charging optimal control problems~\cite{malin,selector}.} Therefore, when designing practical batteries, it is important to consider charging speed along with dendrite suppression.}

\textcolor{newred}{We present a scalable approach for exploration and optimization in the multi-dimensional parameter space of phase-field models, which provides a significant freedom of design through the choice of objective function. In particular, we demonstrate how additional design objectives, such as fast charging, can be incorporated alongside dendrite inhibition to obtain a more balanced solution.} Our approach is based on Bayesian optimization, a machine-learning tool for optimization of black-box functions. We use this tool to find the optimal parameters of a battery cell that maximize an objective function, which is defined based on the electrode/electrolyte interface evolution of the battery cell during an electrodeposition process. \textcolor{newred}{The parameters space can be explored more efficiently in this way, requiring much fewer trials than an exhaustive search~\cite[\S 1]{Garnett}. In particular,} by considering fast charging and dendrite suppression in the objective function, we obtain simple design rules, which reduce dendrite growth or increase the charging speed under a constant voltage, by tuning a few chemical parameters. 

\textcolor{newred}{Phase-field simulations of dendrite growth with fixed parameters and Bayesian optimization of costly functions are established tools in material design and machine learning. The novelty of this work is their integration for multi-objective design in battery systems.} \textcolor{newred}{The proposed approach can identify non-trivial parameter combinations that promote dendrite-free growth under fast-charging conditions. As a result, it simplifies the battery design process by replacing manual trial and error based on lengthy simulations or costly experiments.} 

Our numerical simulations suggest that the interface evolution in a short time interval can provide reliable information on dendrite formation in prolonged charging sessions in some cases. Therefore, by shortening the observation interval and using the onset of dendrite formation, we further reduce the computational effort used in this framework. Moreover, we localize the phase-field equations in both space and time to gain more insight into how each parameter contributes to dendrite growth and charging speed. This localization reduces the system of partial differential equations in a phase-field model to a two-point boundary value problem in a small area of space and time. Using this reduced model, we define a few descriptors for the shapes of the electric and chemical potential fields and the deposition rate variations along the electrode/electrolyte interface. \textcolor{newred}{These descriptors, though approximate, can reveal how certain parameters influence the electrodeposition process without simulating the dendrite evolution even for a short amount of time.} 



\textcolor{newred}{This paper is organized as follows. We introduce a phase-field model and construct a flexible design framework based on Bayesian optimization that can be used to optimize its parameters. We use this framework to optimize the chemical parameters of a lithium-metal battery with respect to dendrite growth and charging speed. The results yield simple rules that predict the influence of several parameters on the charging session. To study these influences more closely, we derive a few descriptors by a local analysis of the interface during electrodeposition. These descriptors are associated with the shapes of electric and chemical potential fields and the electrodeposition rate variations along the electrode/electrolyte interface. We use these descriptors to gain more insight into the influence of different parameters on the charging session and discuss our findings relative to the literature.}

\section{Results}\label{sec:results}

\subsection{Phase-field equations}~\label{sec:phasefield}
Phase-field equations provide a unique way to study the morphological evolution of electrodeposited metals during charging processes, without needing to track the interface evolution~\cite{rev1}. This is realized by using a ``soft'' parameter $\zeta\in[0,1]$ for describing phases, which is $\zeta=1$ for the pure solid phase (electrode) and $\zeta=0$ in the pure liquid phase (electrolyte). In this work, we are interested in the grand potential-based phase-field model provided by Hong and Viswanathan in~\cite{hong2018phase}, which describes the relation between the phase parameter $\zeta$, chemical potential $\mu$ and electric potential $\phi$ in charging half cells, using the following system of partial differential equations:
\begin{align}
\partial_t\zeta&=-L_{\sigma}\bigl(g'-k\nabla^2\zeta\bigr)-L_{\eta}q \label{eqn:zeta}\\
{\partial_\mu c}\,\partial_t\mu&=\nabla \,.\, p\,(\nabla \mu+nF\nabla\phi)-\partial_\zeta c\,\partial_t\zeta \label{eqn:mu}\\
    \nabla\,.\,(\sigma \nabla \phi)&=nFC^{s}_m\partial_t\zeta, \label{eqn:phi}
\end{align}
where \textcolor{newred}{$\nabla.$ and $\nabla^2$ are divergence and Laplace operators respectively, and}
\begin{align}\label{eqn:phasefield_2}
    h(\zeta)&=\zeta^3(6\zeta^2-15\zeta+10)\\
    g(\zeta)&=W\zeta^2(1-\zeta)^2\nonumber\\
    c^{l,s}(\mu)&={\exp\left(\frac{\mu-\epsilon^{l,s}}{RT}\right)}/\left({1+\exp\left(\frac{\mu-\epsilon^{l,s}}{RT}\right)}\right)\nonumber\\
    c_{M^{n+}}(\zeta,\mu)&=c^l(\mu)(1-h(\zeta))\nonumber\\
    q(\zeta,\mu,\phi)&=h'(\zeta)\left(\exp\left(\frac{(1-\alpha)nF}{RT}(\phi-E^{\theta})\right)\right.\nonumber\\
    &\left.-\frac{c_{M^{n+}}(\zeta,\mu)}{c_0}
    \exp\left(-\frac{\alpha nF}{RT}(\phi-E^{\theta})\right)\right)\nonumber\\
    p(\zeta,\mu)&={D^l(1-h(\zeta))c_{M^{n+}}(\zeta,\mu)}/{RT}\nonumber\\
    c(\zeta,\mu)&=c^l(\mu)(1-h(\zeta))+c^s(\mu)h(\zeta){C^s_m}/{C^l_m}\nonumber\\
    \sigma(\zeta)&=\sigma^sh(\zeta)+\sigma_l(1-h(\zeta)).\nonumber
\end{align}
In (\ref{eqn:phasefield_2}), $h$ is an interpolation function for a smooth diffuse interface and $h'$ denotes its derivative, $g$ is a double-well function describing the equilibrium states of solid and liquid phases, $c^s$ and $c^l$ are the molar ratios in the solid and liquid phases, $c_{M^{n+}}$ is the local ion molar ratio, $q$ is the driving force, $c$ is the total concentration, and $\sigma$ is the effective conductivity.

\textcolor{newred}{Equation (\ref{eqn:zeta}) is derived by considering a linear dependence of the phase-ﬁeld evolution on the interfacial free energy part of the driving force, and a nonlinear (exponential) dependence on the thermodynamics driving force, which captures the Butler-Volmer electrochemical reaction kinetics~\cite{chen2015modulation}. Equation (\ref{eqn:mu}) is derived from the mass conservation law~\cite{hong2018phase}
$$
\partial_tC=-\nabla.J,
$$
where $C$ is the volume concentration of the $M$ species (including the $M$ metal and $M^{n+}$ ions) and $J$ is their flux. Equation (\ref{eqn:phi}) is derived from the current density conservation, with a source term $I=nFC^{s}_m\partial_t\zeta$ to account for the charge entering and leaving the system due to the electrochemical reactions~\cite{chen2015modulation}. Note that we assume the electrodeposition proceeds under a constant voltage by placing a constant Dirichlet boundary condition on the electric potential $\phi$ in (\ref{eqn:phi}) throughout the charging session. Hence, while the current rate depends on this voltage, it remains implicit in the model and can vary over time. For more details on the model (\ref{eqn:zeta})-(\ref{eqn:phi}), including its parameters, initial and boundary conditions, and its detailed derivations, see Section~\ref{Sec:Methods} and consult~\cite{hong2018phase,chen2015modulation}.}


This paper aims at finding the parameters that optimize the electrodeposition process modeled by the phase-field equations (\ref{eqn:zeta})-(\ref{eqn:phi}). Several of these parameters are interconnected. For example, conductivity $\sigma^l$ and diffusivity $D^l$ are related through the Nernst–Einstein equation, and interfacial mobility $L_{\sigma}$ and reaction coefficient $L_{\eta}$ are related via the Allen Cahn equation~\cite{LL1,LL2}. We relax these constraints, by assuming the parameters of the phase-field model (\ref{eqn:zeta})-(\ref{eqn:phi}) can be chosen independently. \textcolor{newred}{
In addition, some parameters, such as the diffusion coefficient $D^l$, are not constant in a real system and can vary across time and space during electrodeposition due to, for example, ion depletion near the interface. These parameters are assumed to be constant in the model (\ref{eqn:zeta})-(\ref{eqn:phi}) for tractability reasons. Finally, while certain parameters in the model can be influenced through, for example, material design or synthesis in a conventional way, some parameters are more difficult to tune in a physical system. In this paper, we focus on identifying regions where stable dendrite-free growth is predicted, which can then inform experimental strategies for material development and processing. Hence, the proposed optimization tool is considered a design guide for battery systems, rather than a means to arbitrarily tune physical parameters. 
}


\subsection{\textcolor{newred}{Optimization framework}}\label{sec:BO}
{In this section, we develop a framework to find parameters of the phase-field model (\ref{eqn:zeta})-(\ref{eqn:phi}) that maximize a user-defined objective function. This framework is based on Bayesian optimization and is used to minimize dendrite growth and maximize charging speed in Section~\ref{sec:results}.}

Let $\partial^2u=[u,\partial_xu,\partial_yu,\partial_{xx}u,\textcolor{newred}{\partial_{xy}u},\partial_{yy}u]$ denote the vector of {spatial} partial derivatives of the function $u(x,y,t)$ up to order two. The phase-field equations can be written as the following shorthand nonlinear system of partial differential equations in space $(x,y)$ and time $t$:
\begin{align}    \partial_t\zeta&=f_1(\partial^2\zeta,\mu,\phi;\,\theta)\\
\partial_t\mu&=f_2(\partial^2\zeta,\partial^2\mu,\partial^2\phi;\,\theta)\\
0&=f_3(\partial^2\zeta,\mu,\partial^2\phi;\,\theta),
\end{align}
where the independent variables are omitted for convenience, and $\theta$ is a vector of selected constant parameters related to the chemical properties of the battery. We are interested in finding the optimal parameters $\theta^{\star}\in\Theta$ that result in the optimal plating (in a sense defined by the objective function) after charging the battery under constant voltage from the given initial state
\begin{align}\label{eqn:ini}
    \zeta(x,y,0)=\zeta_0(x,y)\\
    \mu(x,y,0)=\mu_0(x,y)\nonumber
\end{align}
up until $t=t_f$. The set $\Theta$ determines the admissible ranges for each parameter in $\theta$ to prune out solutions that are unreasonable or practically infeasible. 

There is a freedom to choose the objective function based on the design goals. Herein, we are interested in dendrite inhibition and fast charging. Therefore, we choose an objective function that combines two functions corresponding to these two objectives. These functions measure dendrites and the state of charge in a cell.

To minimize dendrite growth in a cell, one first needs to define a function that measures the extent of dendrites. Several different functions have been proposed for this purpose in the literature, such as the distance between the longest dendrite's tip and the deepest valley in the cell (the maximum height), the time it takes dendrites to reach the opposite electrode (short-circuit time), and the relative length of the path that follows the electrode surface from top to bottom (tortuosity)~\cite{mueffect,ahmad}.

\textcolor{newred}{Consider a half cell with the (metal) electrode located at $x=x_0$ and the electrolyte on the opposite side, \emph{i.e.}, $x=x_f$. The length of the half cell is $x_f-x_0$ while the width of the half cell is given by $y_f-y_0$.} In the present work, we use the following dendrite measure based on the order parameter $\zeta$:
\begin{align}\label{eqn:ro}    \rho(t;\,\theta)&=\max_{y\in[y_0,y_f]}\int_{x_0}^{x_f}\zeta(x,y,t)dx\\
&-\min_{y\in[y_0,y_f]}\int_{x_0}^{x_f}\zeta(x,y,t)dx. \nonumber
\end{align}
The dendrite function (\ref{eqn:ro}) generalizes the maximum height measure in \cite{ahmad} to soft phase orders. Namely, if the phase order $\zeta$ is crisp at time $t$, \emph{i.e.}, when
\begin{align*}
    \zeta(x,y,t)=1,&\quad  x\leq x_0(y,t)\\
    \zeta(x,y,t)=0,&\quad  x > x_0(y,t),
\end{align*}
then definition (\ref{eqn:ro}) falls back to the maximum height measure in \cite{ahmad} and $\rho(t)=0$ holds if and only if the surface of the electrode is flat.

{Different dendrite functions were observed to yield} similar results when studying dendrite growth during charging~\cite{mueffect,ahmad}. Nevertheless, we have chosen function (\ref{eqn:ro}) for two main reasons. First, despite tortuosity, it better captures the needle-like dendrites that risk internal short circuits, {a well-known cause of thermal run-away in batteries~\cite[\S 6.1.2.1]{berg2015batteries}.} Second, despite the short-circuit time, it can be computed using the current order parameter value with no need to simulate the charging process until short-circuit.

To maximize the charging speed, we maximize the state of charge at the end of the charging session $t_f$. The state of charge at time $t$ can be estimated by measuring the surface of the electrode relative to the total area of the cell as follows
\begin{equation}\label{eqn:SOC}
    S(t;\theta)=\frac{\int_{y_0}^{y_f}\int_{x_0}^{x_f} \zeta(x,y,t) dxdy}{2(x_f-x_0)(y_f-y_0)},
\end{equation}
where the denominator is the area of the cell (double the half-cell area).

Note that since the order parameter $\zeta$ evolves differently for different parameter values in (\ref{eqn:zeta})-(\ref{eqn:phi}), both $\rho$ and $S$ {are functions} of the selected parameters $\theta$ in (\ref{eqn:ro}) and (\ref{eqn:SOC}). Therefore, to inhibit dendrite growth and accelerate charging speed, one may look for the optimal parameters $\theta^{\star}$ \textcolor{newred}{that \emph{minimize} the dendrite function $\rho(t;\,\theta)$ but \emph{maximize} the state of charge $S(t;\theta)$. Hence, settling for a trade-off, one can 
maximize} an averaged objective function as follows:
\begin{equation}\label{eqn:cost}
    C(t,\theta)=-\lambda \rho(t;\theta)+(1-\lambda)\bar{S}(t;\theta),
\end{equation}
where we have scaled the state of charge as $\bar{S}(t;\theta)=2(x_f-x_0)S(t;\theta)$ to make sure $\bar{S}(t;\theta)$ and $\rho(t;\theta)$ are within the same range
\begin{align*}
    \bar{S}(t;\theta),
    \rho(t;\theta)\in[0,x_f-x_0].
\end{align*}
In (\ref{eqn:cost}), $\lambda\in[0,1]$ specifies the trade-off between the two objectives. To speed up charging, we can choose a smaller value for $\lambda$, while to inhibit dendrites, we can increase $\lambda$. Therefore, we consider the following optimization problem:

\begin{equation}
    \begin{array}[c]{rll}
    \underset{\theta\in\Theta}{\text{maximize}} & C(t_f;\,\theta)\\
    \mbox{subject to}
    & \zeta,\mu,\phi\textnormal{ satisfy (\ref{eqn:zeta})-(\ref{eqn:phi})}.
    \end{array} \label{eqn:opt}
\end{equation}
Evaluating the objective function in (\ref{eqn:opt}) at a given point $\theta$ requires simulating the phase-field equations (\ref{eqn:zeta})-(\ref{eqn:phi}) for $t\in[0,t_f]$ to obtain $\zeta(x,y,t_f)$, which is plugged in (\ref{eqn:ro}) and (\ref{eqn:SOC}) to compute the dendrite and state-of-charge functions.
This process requires a significant amount of time and computation power. {In addition, no explicit expression is known for the objective function in (\ref{eqn:opt}) and there is no efficient method available for estimating its gradient, rendering most traditional optimization paradigms inapplicable. Therefore, we consider the objective function (\ref{eqn:opt}) as a black box and use Bayesian optimization to solve the problem~(\ref{eqn:opt}). The remarkable sample efficiency of Bayesian optimization further motivates this choice among other global optimization routines, because of the costliness of function evaluations in (\ref{eqn:opt})~\cite[\S 1]{Garnett}.}

\subsection{\textcolor{newred}{Bayesian optimization}}
\textcolor{newred}{
Bayesian optimization is a global optimization routine used to optimize black-box functions that are costly to evaluate, like $C(t_f;\theta)$ in (\ref{eqn:opt}). In this method, the objective function is replaced with a surrogate model that is improved iteratively by evaluating the objective function at selected points. This surrogate model, which is optimized instead of the original objective function, is a Gaussian process. A Gaussian process is a random functional $G(\theta)$ whose values at any finite collection of points in $\Theta$ make a joint Gaussian (normal) distribution. These distributions have a long track of success in modeling complex multi-objective problems~\cite{Garnett,rev2}. Since Gaussian distributions are completely determined by their first and second moments, a Gaussian process $G_i(\theta)\sim \mathcal{GP}(m_i,K_i)$ is also defined without ambiguity by its mean $m_i$ and kernel function $K_i$ as follows:
\begin{align}
m_i(\theta)&=\mathbb{E}(G_i(\theta))\nonumber\\
K_i(\theta,\theta')&=\operatorname{Cov}\bigl(G_i(\theta),G_i(\theta')\bigr). \label{mean&kernel}
\end{align}
Bayesian optimization starts at iteration $i=0$ with an initial model (called the prior), by choosing (\ref{mean&kernel}) based on an initial belief that reflects the domain knowledge. For example, if $\theta\in\mathbb{R}$ is the applied voltage, one can choose the prior such that $m_0(\theta)$ increases with $\theta$, as a higher voltage is often associated with more dendrite growth and faster charging. However, more frequently, the prior $G_i$ is chosen automatically, by evaluating $C(t_f;\theta)$ at a few (often randomly selected) points in $\Theta$ and finding the functions $m_0$, $K_0$ that maximize the likelihood
$$
\mathbb{P}\bigl\lbrace\mathcal{D}_0\,\vert\, m_0,K_0\bigr\rbrace,
$$
where $\mathcal{D}_0$ contains the data from the initial function evaluations. After setting a prior, an algorithm chooses a point $\theta_i\in\Theta$, evaluates the objective function at this point, and uses Bayesian inference to update the current model $G_i$ according to this new information $\mathcal{D}_i$, as follows:
\begin{equation}\label{eqn:updateG}
    \mathbb{P}\bigl\lbrace G_{i+1}(\theta)=\omega\bigr\rbrace=
\mathbb{P}\bigl\lbrace G_{i}(\theta)=\omega\,\vert\, \mathcal{D}_i\bigr\rbrace,
\end{equation}
where $\omega\in\mathbb{R}$. The updated model is also a Gaussian process, \emph{i.e.}, $G_{i+1}(\theta)\sim\mathcal{GP}(m_{i+1},K_{i+1})$, whose mean and kernel functions are easily computed from $m_i$, $K_i$, and the new data $\mathcal{D}_i$. This process is repeated for $i=1,2,\dots$ until a stopping criterion is reached.}

\textcolor{newred}{In the above procedure, the evaluation points $\theta_i$ are chosen to explore regions where the objective function is expected to be large as well as high-uncertainty regions. This is achieved by first encoding exploration preferences within the feasible set $\Theta$ using an acquisition function $A(\theta\,\vert\, \mathcal{D}_0,\mathcal{D}_1,\dots,\mathcal{D}_i)$, and then choosing the next evaluation point as one that maximizes this preference in each iteration, \emph{i.e.},
\begin{equation}\label{eqn:updatetheta}
    \theta_{i+1}\in\arg\max A(\theta\,\vert\, \mathcal{D}_0,\mathcal{D}_1,\dots,\mathcal{D}_i).
\end{equation}   
Therefore, a Bayesian optimization algorithm improves our estimate of the objective function by updating the surrogate model $G_i$ as (\ref{eqn:updateG}) and finds the regions where the objective function is maximized by updating the evaluation point $\theta_i$ as (\ref{eqn:updatetheta}). Details about the specific Bayesian optimization algorithm used in this paper can be found in Section~\ref{Sec:Methods}.
}

\textcolor{newred}{\subsection{Application to lithium-metal batteries}}
We apply the Bayesian optimization framework developed above \textcolor{newred}{to find the optimal values for the electronic conductivity of the solid phase $\sigma^s$, ionic conductivity of the liquid phase $\sigma^l$, site density in electrolyte $C^l_m$, diffusion coefficient in electrolyte $D^l$, interfacial mobility $L_{\sigma}$, electrochemical reaction kinetic coefficient $L_\eta$, interface tension $\gamma$, and interfacial thickness $\delta$} in a lithium-metal half cell. The feasible intervals used for these parameters are listed in Table~\ref{tab:ranges}. These intervals are derived from perturbing the nominal values in \cite{Rebeca} by a maximum of $50\%$ and include the parameter values used in various studies with different electrodes and electrolytes~\cite{zinc3,Sodium,hong2018phase,chen2015modulation,ahmad,zinc1,L1,Mg,Li2}. The rest of the parameters in (\ref{eqn:zeta})-(\ref{eqn:phi}) are chosen according to \cite{Rebeca}. To make the simulations more realistic, we use an electrode with an initially rough surface, as no electrode has a perfectly flat surface in practice. This initial roughness of the electrode surface also makes dendrites grow faster, which can help to save computation power by choosing a smaller horizon length $t_f$ in (\ref{eqn:opt}).


\begin{table*}[t]
\begin{center}
\begin{tabular}{|l l l l l l|} 
 \hline
 Parameter & notation & unit & value range &
 $\lambda=1$ & $\lambda=0$ \\ [0.5ex] 
 \hline\hline
 Electronic conductivity of the solid phase & $\sigma^s$ & [S/m] & $10^{6}\times [0.5, 1.5]$ & $1.5\times 10^{6}$ & $0.5\times 10^6$\\ 
 Ionic conductivity of the liquid phase & $\sigma^l$ & [S/m] & $[0.5950, 1.7850]$ & $0.5950$ & $0.6866$\\
  Site density in electrolyte & $C^l_m$ & [mol/$\textnormal{m}^3$] & $10^{4} [0.9261, 2.7782]$ & $0.9261 \times 10^{4}$ & $2.7782 \times 10^4$\\
  Diffusion coefficient of ions $M^{n+}$ in electrolyte & $D^l$ & [$\textnormal{m}^2$/s] & $10^{-9} [0.1590 , 0.4769]$ & $0.1590 \times 10^{-9}$ & $0.4769\times10^{-9}$\\
  Interfacial mobility & $L_\sigma$ & [$\textnormal{m}^3$/Js] & $10^{-5} [0.1250,  0.3750]$ & $0.3750 \times 10^{-5}$ & $0.3750 \times 10^{-5}$\\
 Electrochemical reaction kinetic coefficient & $L_\eta$ & [$\textnormal{s}^{-1}$] & $[0.0005,  0.0015]$ & $0.0005$ & $0.0015$\\
  \textcolor{newred}{Interface tension} & \textcolor{newred}{$\gamma$} & \textcolor{newred}{[J/$\textnormal{m}^2$]} & \textcolor{newred}{$[0.45,  0.55]$} & \textcolor{newred}{$0.45$} & \textcolor{newred}{$0.55$} \\
\textcolor{newred}{Interfacial thickness} & \textcolor{newred}{$\delta$} & \textcolor{newred}{[m]} & \textcolor{newred}{$10^{-6}[0.9,  1.1]$} & \textcolor{newred}{$1.1\times 10^{-6}$} & \textcolor{newred}{$1.1\times 10^{-6}$}\\
 \hline
\end{tabular}
\caption{The optimal parameters that maximize the objective function (\ref{eqn:cost}) with different values of $\lambda$ found by Bayesian optimization. The parameters in the column $\lambda=0$ maximize the charging speed, while the parameters in the column $\lambda=1$ inhibit dendrite growth.}
\label{tab:ranges}
\end{center}
\end{table*}

\begin{table}[t]
\begin{center}
\begin{tabular}{|l l |} 
 \hline
 Parameter & Guideline \\ [0.5ex] 
 \hline\hline
 Electronic conductivity of the solid phase $\sigma^s$ & Increase \\ 
 Ionic conductivity of the liquid phase $\sigma^l$ & Decrease \\
  Site density in electrolyte $C^l_m$ & Decrease \\
  Diffusion coefficient of ions $M^{n+}$ in electrolyte $D^l$ & Decrease \\
  Interfacial mobility $L_\sigma$ & Increase \\
 Electrochemical reaction kinetic coefficient $L_\eta$ & Decrease \\
  \textcolor{newred}{Interface tension} \textcolor{newred}{$\gamma$} & \textcolor{newred}{Decrease} \\
\textcolor{newred}{Interfacial thickness} \textcolor{newred}{$\delta$} & \textcolor{newred}{Increase} \\
 \hline
\end{tabular}
\caption{
Summary of Table~\ref{tab:ranges} for $\lambda=1$. Guideline on choosing the parameter values for \emph{minimal dendrite growth} in the lithium-metal half cell considered in Section~\ref{sec:results}.} 
\label{tab:ruleofthumb}
\end{center}
\end{table}

\begin{table}[t]
\begin{center}
\begin{tabular}{|l l |} 
 \hline
 Parameter & Guideline \\ [0.5ex] 
 \hline\hline
 Electronic conductivity of the solid phase $\sigma^s$ & Decrease \\ 
 Ionic conductivity of the liquid phase $\sigma^l$ & Increase \\
  Site density in electrolyte $C^l_m$ & Increase \\
  Diffusion coefficient of ions $M^{n+}$ in electrolyte $D^l$ & Increase \\
  Interfacial mobility $L_\sigma$ & Increase \\
 Electrochemical reaction kinetic coefficient $L_\eta$ & Increase \\
 \textcolor{newred}{Interface tension} \textcolor{newred}{$\gamma$} & \textcolor{newred}{Increase} \\
\textcolor{newred}{Interfacial thickness} \textcolor{newred}{$\delta$} & \textcolor{newred}{Increase} \\
 \hline
\end{tabular}
\caption{
Summary of Table~\ref{tab:ranges} for $\lambda=0$. Guideline on choosing the parameter values for \emph{maximum charging speed} in the lithium-metal half cell considered in Section~\ref{sec:results}.} 
\label{tab:ruleofthumb2}
\end{center}
\end{table}

\begin{figure*}[t]
     \centering
    \begin{subfigure}[b]{0.45\textwidth}
         \centering    \includegraphics[width=1\linewidth]{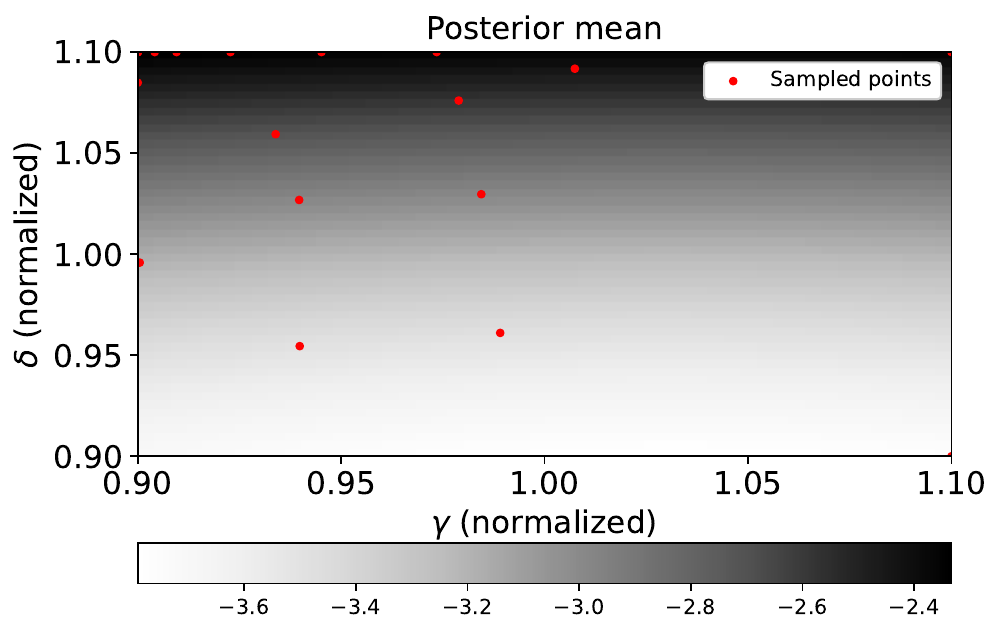}
	    \caption{\textcolor{newred}{The mean of the Gaussian process that predicts the objective function (\ref{eqn:cost}). The dark shade represents the regions where the model expects a \emph{higher} value for the objective function, and hence, less dendrite growth.}}
       \label{fig:mean}
    \end{subfigure}\hspace{5mm}
    \begin{subfigure}[b]{0.45\textwidth}
        \centering
     \includegraphics[width=1\linewidth]{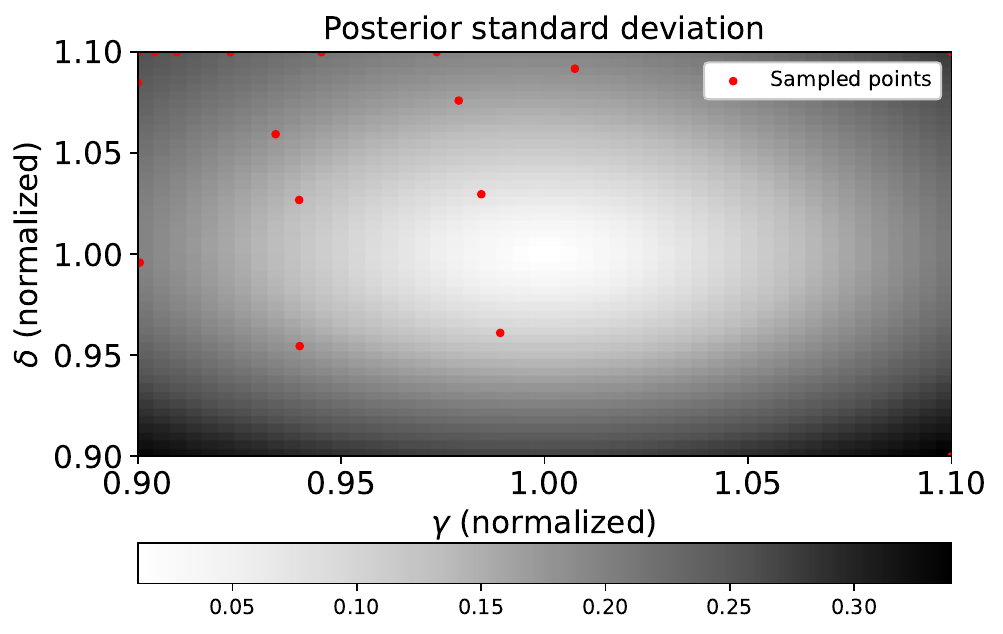}
	    \caption{\textcolor{newred}{Standard deviation of the Gaussian process that predicts the objective function (\ref{eqn:cost}). The dark shade represents the regions where the model is more \emph{uncertain} about the objective function.}}
         \label{fig:std}
     \end{subfigure}
        \caption{\textcolor{newred}{The posterior distribution predicted by the Bayesian optimization framework  after $20$ samples, where $\theta=[\gamma,\delta]^\top$ is optimized. According to Figure~\ref{fig:std}, the model is less certain in the regions where $\delta$ is smaller. However, as the location of sampled points shows, the algorithm has favored exploring the opposite region due to its higher expected rewards in Figure~\ref{fig:mean}.}}
        \label{fig:posterior}
\end{figure*}


\begin{figure*}[t]
     \centering
     \begin{subfigure}[b]{0.45\textwidth}
        \centering
        \includegraphics[width=1\linewidth]{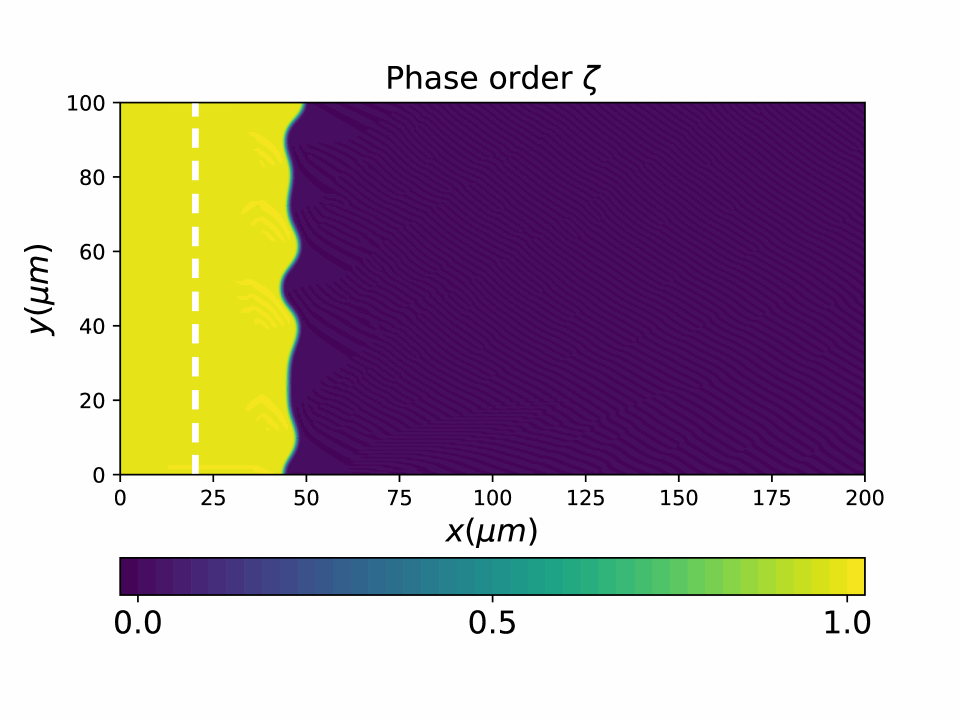}
	    \caption{Dendrite-prone deposition. The half cell with parameters chosen from \cite{Rebeca} develops dendrites {and reaches the state of charge $0.12$ in $60$ (s)}.}
         \label{fig:slow_full}
     \end{subfigure}
    \begin{subfigure}[b]{0.45\textwidth}
         \centering
        \includegraphics[width=1\linewidth]{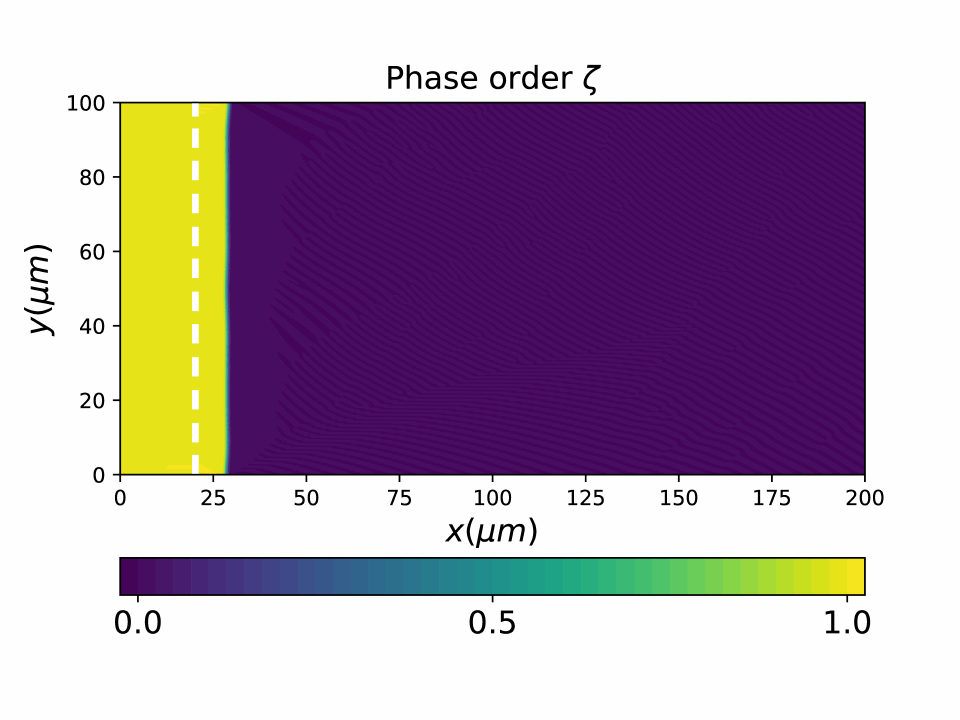}
	    \caption{Dendrite-free deposition. The half cell with parameters chosen according to Table~\ref{tab:ruleofthumb} forms no dendrites and {reaches the state of charge $0.07$ in $60$ (s)}.}
       \label{fig:slow_free}
    \end{subfigure}
        \caption{Two half cells charged with the same constant voltage $-0.45$ (v) and the same amount of time $60$ (s). The dashed line shows the initial interface at the beginning of the charging process.}
        \label{fig:slow}
\end{figure*}

\begin{figure}
	\begin{center}
    \includegraphics[width=1\linewidth]{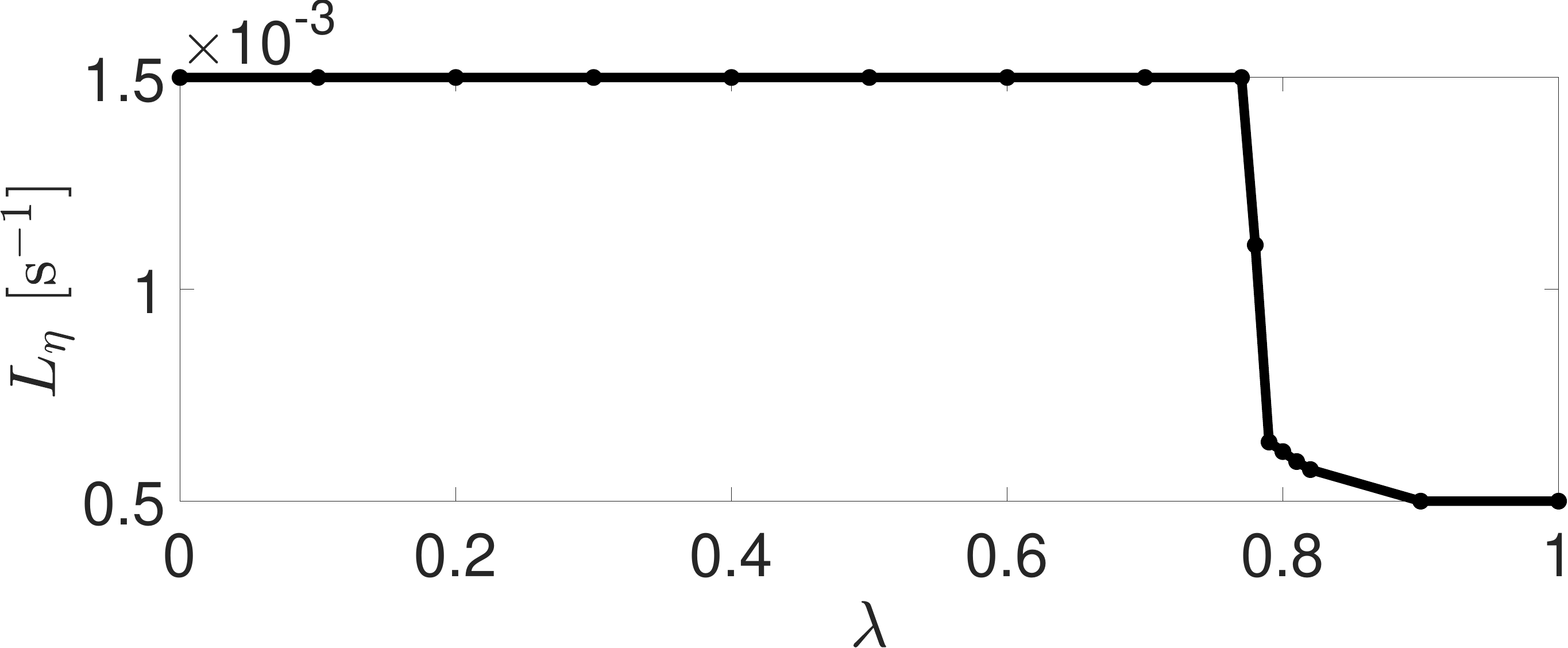}  
	\caption{\textcolor{newred}{The optimal electrochemical reaction kinetic coefficient $L_{\eta}$ as a function of the trade-off parameter $\lambda$ in (\ref{eqn:opt}). Despite the monotonic relations reported in Tables~\ref{tab:ruleofthumb} and \ref{tab:ruleofthumb2}, the relationship between $L_{\eta}$ and the objective function is no longer monotonic here.}}
	\label{fig:lamplot}
	\end{center}
\end{figure}

To only minimize dendrite growth, we choose $\lambda=1$ in the objective function in (\ref{eqn:opt}). As a first experiment, we solve the optimization problem (\ref{eqn:opt}) with a short time horizon $t_f=0.04$ (s) once for each parameter $\theta\in\mathbb{R}$ in Table~\ref{tab:ranges}. The rest of the parameters are kept constant at the center of their intervals. All the parameters are found to be optimal at their extreme values as shown under the column $\lambda=1$ of Table~\ref{tab:ranges}. \textcolor{newred}{A simple rule of thumb can summarize the results of this experiment for tuning the battery parameters to inhibit dendrites, as shown in Table~\ref{tab:ruleofthumb}. One can gain more insight by evaluating the posterior distribution obtained within the Bayesian optimization algorithm. For example, this distribution is shown in Figure~\ref{fig:posterior}, when interfacial thickness $\delta$ and interfacial tension $\gamma$ are considered as the optimization variables. Figure~\ref{fig:posterior} indicates that increasing the interfacial thickness is probably more effective than reducing the interfacial tension in maximizing the objective function (\ref{eqn:cost}), and hence, minimizing dendrite growth. This kind of information can help prioritize the parameters in battery design when there are more restrictions for tuning.}

In the second experiment, we increase the horizon length in the first experiment by $2.5$ times, \emph{i.e.}, $t_f=0.1$ (s) and optimize all the first four parameters in Table~\ref{tab:ranges} simultaneously ($\theta\in\mathbb{R}^4$ ). The results of this experiment coincide with those of the first experiment perfectly. This suggests that the optimal solution to (\ref{eqn:opt}) is not sensitive to $t_f$. \textcolor{newred}{This temporal independence of the optimal parameters is also observed in \cite[Fig.7]{zinc3} and \cite[Fig.5]{Mg} for the applied voltage, in~\cite[Fig.11]{zinc3} for the exchange current density, in~\cite[Fig.6c]{zinc1} for flowing velocity, and in~\cite[Fig.12]{L1} for interfacial thickness. In these papers, a parameter that gives the least dendrite growth in a short period of time also results in the smallest dendrites in longer simulations. We note, however, that this observation does not apply to situations where the system behavior changes dramatically during the charging session. This happens when a diffusion-controlled regime takes over the system after the current peaks in \cite[Fig.1c]{LL1}, for example. Nevertheless, when there is no major change in system dynamics, the optimal parameters are less sensitive to time, in which case,} one may choose a smaller value for $t_f$ to make the function evaluation steps less costly and reduce the computation burden in solving the optimization problem (\ref{eqn:opt}) significantly.



\textcolor{newred}{To validate the above results,} we consider charging a half cell {at} the constant voltage $-0.45$ (v) using two different values for the parameters $\sigma^s$, $\sigma^l$, $D^l$, $C^l_m$ and compare the results in Figure~\ref{fig:slow}. When these parameters are chosen as the midpoints of their respective intervals in Table~\ref{tab:ranges}, dendrites grow (Figure~\ref{fig:slow_full}). However, when these parameters were chosen based on the guidelines of Table ~\ref{tab:ruleofthumb} (with the exact values given in column $\lambda=1$ of Table~\ref{tab:ranges}), no dendrites grow in this experiment (see Figure~\ref{fig:slow_free}). This observation confirms the results of Table~\ref{tab:ruleofthumb}. {However, it also has two important indications. First, it suggests that the parameters that minimize dendrite growth within a short time interval $t_f=0.04$ (s) can also inhibit dendrites further in the future $t_f=60$ (s). Second, it shows dendrite inhibition can be at odds with fast charging. Because the amount of charge delivered in the two half cells during the same time interval of $60$ (s) is strikingly different in Figure~\ref{fig:slow}. The cell state of charge at the end of the charging session in Figure~\ref{fig:slow_free} is estimated to be $0.072$, that is, around $37\%$ less than that in Figure~\ref{fig:slow_full}.} Therefore, choosing extreme values for dendrite inhibition can lead to conservative solutions that slows the charging process.


\textcolor{newred}{To only maximize the charging speed, we choose $\lambda=0$ in the objective function~(\ref{eqn:cost}).} Bayesian optimization was used to solve this optimization problem with $t_f=0.04$ (s). The results are reported under the column $\lambda=0$ in Table~\ref{tab:ranges} and summarized as a simple rule in Table~\ref{tab:ruleofthumb2}. Again, all the parameters are found to be optimal at their extreme values, however, on the opposite sides of their spectrums, \textcolor{newred}{except interfacial mobility and thickness}. This agrees with the observation in Figure~\ref{fig:slow} where fast charging and minimal dendrite growth are at odds.


\begin{figure*}[t]
     \centering
     
     \begin{subfigure}[b]{0.45\textwidth}
        \centering
        \includegraphics[width=1\linewidth]{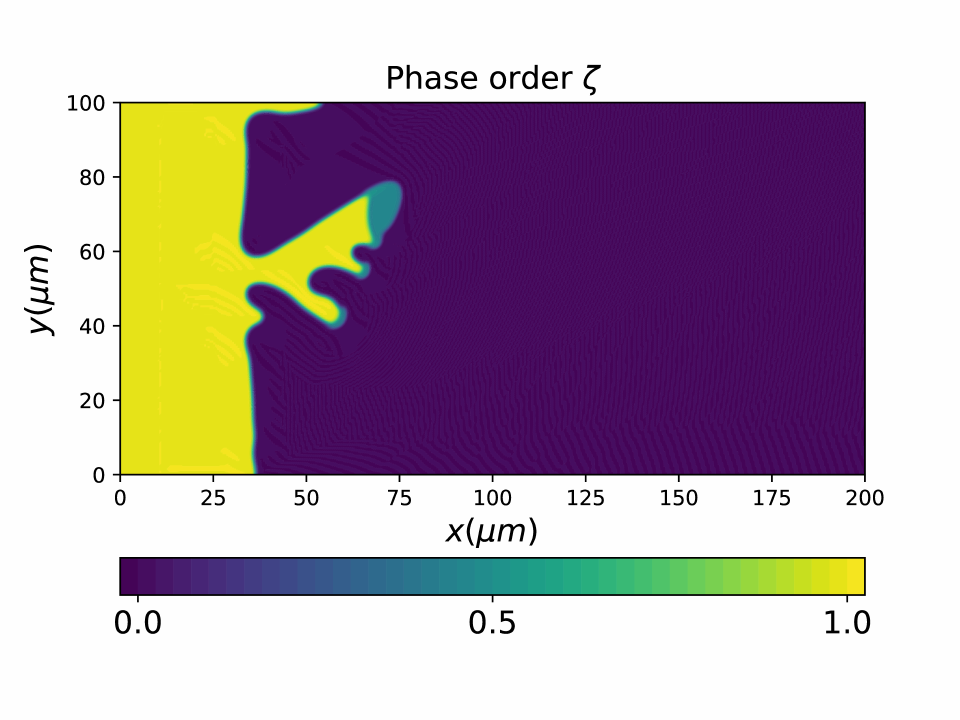}
	    \caption{Dendrite-prone deposition. The half cell with $L_\sigma=0.4\times 10^{-6}$ ($m^3/Js$) develops dendrites {and reaches the state of charge $0.10$ in $110$ (s)}.}
         \label{fig:LsigmaB}
     \end{subfigure}
    \begin{subfigure}[b]{0.45\textwidth}
         \centering
        \includegraphics[width=1\linewidth]{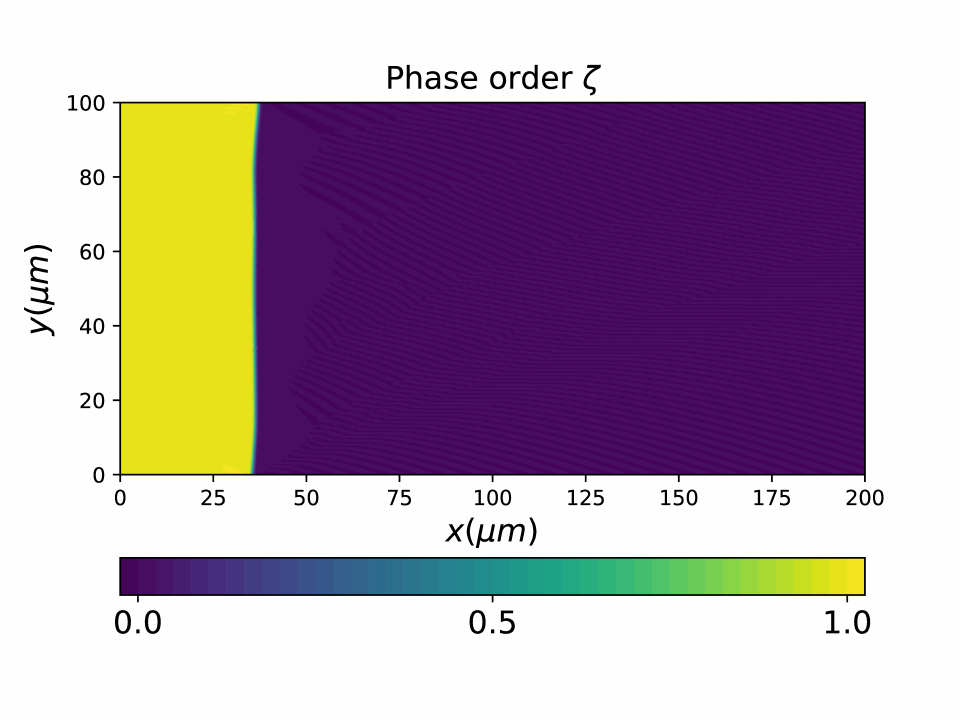}
	    \caption{Dendrite-free deposition. The half cell with $L_\sigma=8\times 10^{-6}$ ($m^3/Js$) {reaches the state of charge $ 0.09$ in $110$ (s) with no visible dendrites}.}
       \label{fig:LsigmaG}
    \end{subfigure}
     \vskip\baselineskip
     \begin{subfigure}[b]{0.45\textwidth}
        \centering
        \includegraphics[width=1\linewidth]{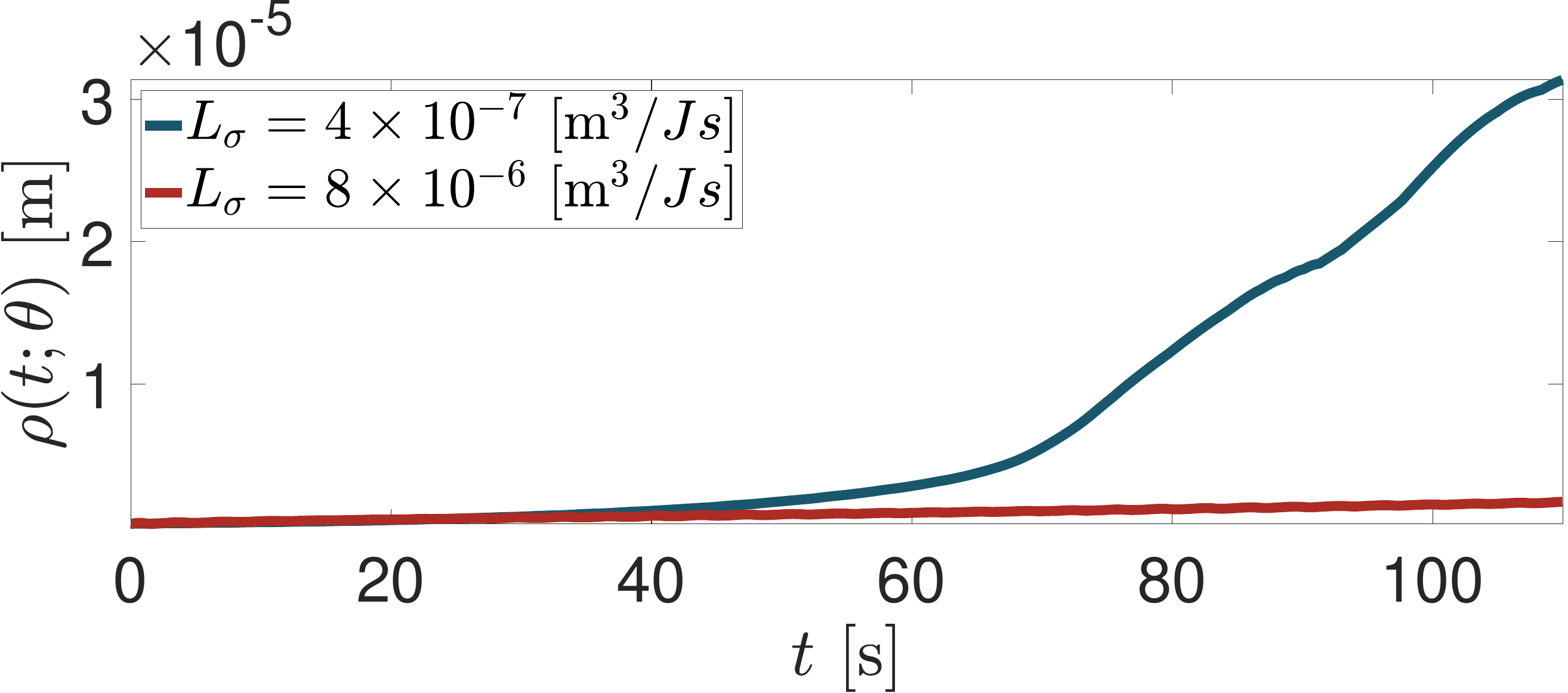}
	    \caption{The roughness of the electrode surface (\ref{eqn:ro}) over time.}
         \label{fig:roughness}
     \end{subfigure}
    \begin{subfigure}[b]{0.45\textwidth}
         \centering
        \includegraphics[width=1\linewidth]{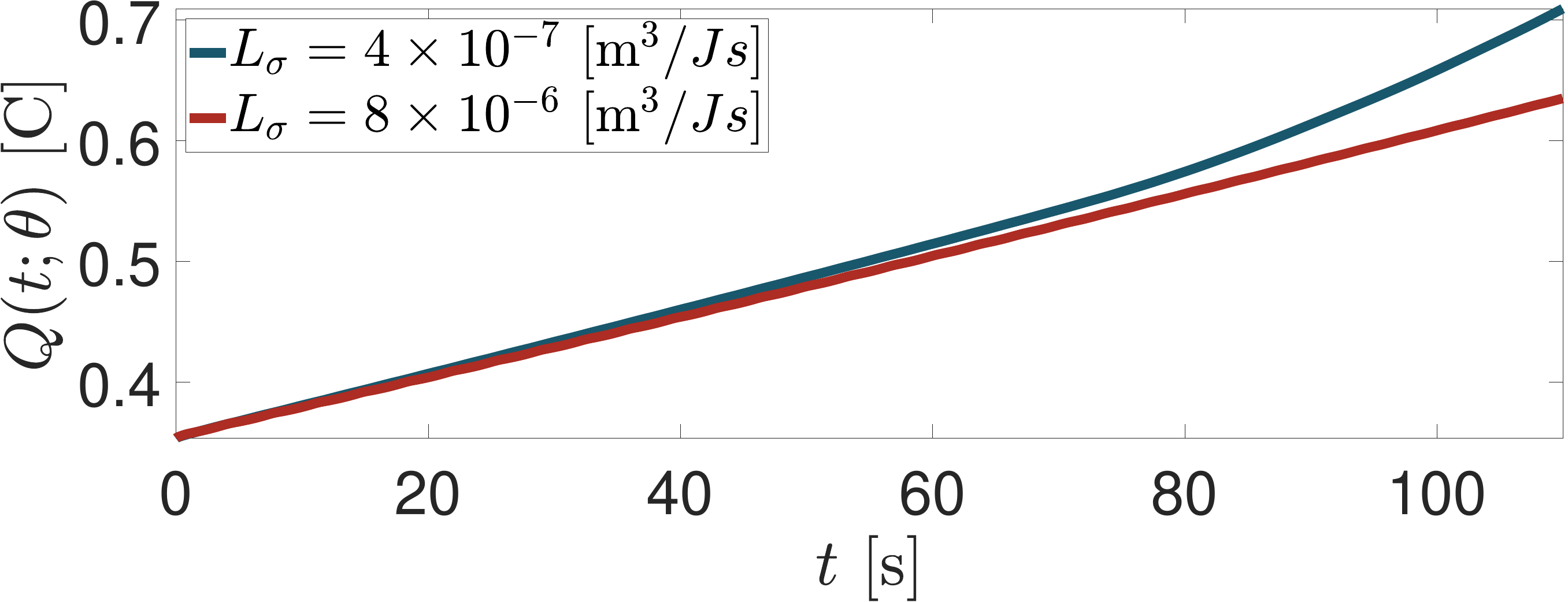}
	    \caption{The delivered charge (\ref{eqn:Q}) over time.}
       \label{fig:charge}
    \end{subfigure}
     
        \caption{Two half cells charged with the same constant voltage $-0.45$ (v) and the same amount of time $110$ (s). The only different parameter between these two half cells is the interfacial mobility $L_\sigma$.}
        \label{fig:Lsigma}
\end{figure*}

One can settle for a middle ground between fast charging and dendrite inhibition by tuning the objective function in the optimization problem~(\ref{eqn:opt}) using $\lambda\in(0,1)$. By using a smaller $\lambda$, the parameters that speed up charging are favored, whereas a larger $\lambda$ would lead to parameters that inhibit dendrite growth. \textcolor{newred}{To examine this effect, we use the optimization framework in Section~\ref{sec:BO} to find the reaction constant $L_{\eta}$ that maximizes the objective function~(\ref{eqn:cost}) with $\lambda\in(0,1)$ and $t_f=0.04$ (s). As shown in Figure~\ref{fig:lamplot}, the optimal reaction rate decreases as dendrite suppression is increasingly favored over fast charging. This result indicates a strong trade-off between the conflicting objectives of dendrite inhibition and fast charging.}

A curious exception to this trade-off is the interfacial mobility $L_\sigma$. This parameter should be maximized to achieve both fast charging and minimum dendrite growth, according to Tables~\ref{tab:ranges}--\ref{tab:ruleofthumb2}. To examine how this parameter affects the charging performance, we consider two half cells with different interfacial mobilities and otherwise equal parameters according to \cite{github}. These half cells are charged with the constant voltage $-0.45$ (v) for $110$ (s). We simulate the dendrite growth in both half cells and compare the results in Figure~\ref{fig:Lsigma}. As shown in Figure~\ref{fig:LsigmaB}, dendrites are only formed in the half cell with the smaller $L_\sigma$. The roughness of the electrode surface (\ref{eqn:ro}) grows more rapidly in this half cell (see Figure~\ref{fig:roughness}). Nevertheless, in both half cells, the electrode surface (excluding the dendrite) is approximately located at $35$~($\mu$m) on the horizontal axis, which indicates a similar charging speed between the two experiments (see Figures~\ref{fig:LsigmaB} and \ref{fig:LsigmaG}). An alternative way to compare the charging speeds between the two experiments is to estimate the delivered electric charge in each half cell up to time $t$ as follows
\begin{equation}\label{eqn:Q}
 Q(t;\,\theta)=nFC^s_m\int_{y_0}^{y_f}\int_{x_0}^{x_f}
 h\left(\zeta(x,y,t)\right)dxdy.
\end{equation}
The above function grows slightly faster in the half cell with the smaller $L_\sigma$. However, the charge delivered in the other half cell is not far behind, growing almost linearly (see Figure~\ref{fig:charge}). Note that both functions (\ref{eqn:SOC}) and (\ref{eqn:Q}) are based on the total area occupied by the solid phase, including the dendrite. At the end of the experiment, the state of charge (\ref{eqn:SOC}) in Figure~\ref{fig:LsigmaG} is $10.6\%$ less than the state of charge in Figure~\ref{fig:LsigmaB}. In comparison, recall that when dendrites were inhibited by tuning the parameters $\sigma^s$, $\sigma^l$, $D^l$, $C^l_m$ instead of $L_\sigma$, the state of charge in Figure~\ref{fig:slow_free} was $37\%$ less than the state of charge in Figure~\ref{fig:slow_full}. Therefore, we conclude that increasing $L_\sigma$ inhibits dendrites more effectively with much less impact on the charging speed.

\subsection{Local analysis of the interface}\label{Sec:Potential}
In Section~\ref{sec:results}, we identified the influence of different chemical parameters on the speed and dendrite formation in a charging process. In this section, we discuss \emph{how} these parameters influence the charging session the way they do. To simplify our discussions, we also quantify several key descriptors for the local behavior of the electrode/electrolyte interface during the electrodeposition process. Analyzing these descriptors instead of the full model (\ref{eqn:zeta})-(\ref{eqn:phi}) helps to understand the mechanism by which different parameters affect the charging process. 

{We begin with approximating the phase-field model (\ref{eqn:zeta})-(\ref{eqn:phi}) across the phase transition interface in an infinitesimal time interval, a process we call \emph{localization}. By restricting space and time in this way, the system of partial differential equations is replaced by an ordinary differential equation around the phase transition interface. This simplified local model provides four descriptors (Table~\ref{tab:drule}) for the shapes of the electric and chemical potential fields and the deposition rate variations along the electrode surface. These descriptors are functions of the battery parameters and provide valuable information on how each parameter contributes to dendrite growth and charging speed.}

\begin{figure}
     \centering
     \begin{subfigure}[b]{0.22\textwidth}
        \centering
        \includegraphics[width=1\linewidth]{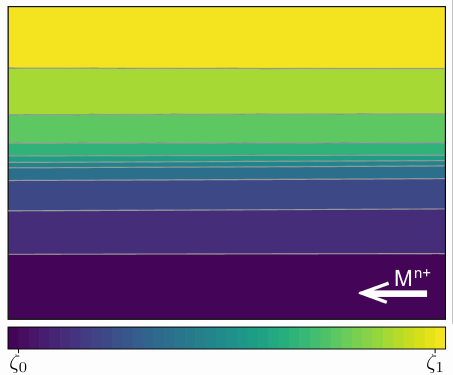}
	    \caption{$z=y$.}
         \label{fig:box_y}
     \end{subfigure}
    \begin{subfigure}[b]{0.22\textwidth}
         \centering
        \includegraphics[width=1\linewidth]{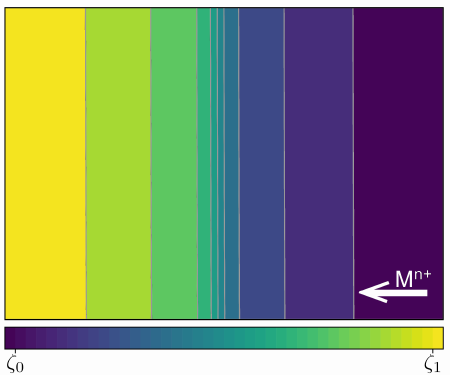}
	    \caption{$z=x$.}
       \label{fig:box_x}
    \end{subfigure}
        \caption{Level sets of the order parameter $\zeta$ in a small box around the electrode/electrolyte transition zone ($\zeta_0<\zeta_1$). The white arrow indicates the charging direction.}
        \label{fig:box}
\end{figure}

Consider a half cell where the negative electrode made of $M$ metal is located on the left and the electrolyte is on the right. During the charging process, the electrochemical potential governs the transport of $M^{n+}$ ions through the electrolyte and their adsorption on the anode. Hence, to understand the effects of different chemical and physical parameters on electrodeposition and dendrite growth, we study their effects on the electric and chemical potentials. To simplify the process, we consider the phase-field equations (\ref{eqn:zeta})-(\ref{eqn:phi}) in a small box within the electrode/electrolyte interface where the phase order parameter varies from $\zeta=\zeta_0$ to $\zeta=\zeta_1$ (see Figure~\ref{fig:box}) where
$$
0<\zeta_0<\zeta_1<1.
$$
Such a local analysis is useful as the electrode/electrolyte interface is where most of the limiting processes occur in a charging cell~\cite[\S1.5]{berg2015batteries}. We assume that the level sets of $\zeta$, $\mu$ and $\phi$ are parallel in this box and
\begin{align}\label{eqn:ass:zeta,mu}
    \zeta&=\bigl(\tanh{\beta(z-z_0)}+1\bigr)/2\\
    \mu&=\mu_s\zeta,\nonumber
\end{align}
where $z=y$ in case {the surface normal inside the box is orthogonal to the charging direction (Figure~\ref{fig:box_y}) and $z=x$ when it is parallel to it (Figure~\ref{fig:box_x}).} In (\ref{eqn:ass:zeta,mu}), $z_0$ specifies the transition interface location, $\beta>0$ determines the sharpness of phase transition, and $\mu_s<0$ is the chemical potential limit in the solid phase. {Functions (\ref{eqn:ass:zeta,mu}) are a smooth approximation of step functions used in the literature in different contexts, such as for representing order parameters and electric potentials~\cite{Rebeca}. Although assuming} parallel level sets may look restrictive, it typically holds inside a small enough box on the tip and sides of grown dendrites (see for example~\cite{Rebeca}). These assumptions simplify the phase-field equations significantly. In particular, substituting (\ref{eqn:ass:zeta,mu}) and (\ref{eqn:zeta}) in the \textcolor{newred}{conduction} equation (\ref{eqn:phi}) results in a second-order \emph{ordinary} differential equation in $\phi$, which after changing the independent variable from $z$ to $\zeta$, takes the form

\begin{equation}\label{eqn:phi_ODE}
    a_2(\zeta)\phi''+a_1(\zeta)\phi'=a_0(\zeta,\phi),
\end{equation}
with the two-point boundary conditions
\begin{equation}\label{eqn:2pbc}
    \phi(\zeta_0)=\phi_0,\;\phi(\zeta_1)=\phi_1
\end{equation}
and coefficient functions
\begin{align}\label{eqn:coeffuncs}
    a_2(\zeta)&=\sigma\zeta(1-\zeta)\nonumber\\
    a_1(\zeta)&=30(\sigma^s-\sigma^l)\zeta^3(1-\zeta)^3-\sigma(2\zeta-1)\nonumber\\
    a_0(\zeta,\phi)&=\frac{nFC^s_m}{2\beta^2}\bigl(L\sigma(2\zeta-1)(W-2\beta^2k) \nonumber\\   
    &-L_{\eta}q(\zeta,\mu_s\zeta,\phi)/2\zeta(1-\zeta)\bigr).
\end{align}
Note that the phase-field equations are symmetrical in the coordinates $x$ and $y$ (except the initial and boundary conditions). Therefore, the above equations are valid for both boxes in Figure~\ref{fig:box} independent of the choice $z\in\lbrace x,y\rbrace$.

Equations (\ref{eqn:phi_ODE})-(\ref{eqn:coeffuncs}) specify a two-point boundary value problem that can be solved by, \emph{e.g.}, collocation and shooting methods to obtain $\phi=\phi(\zeta)$. This solution determines the electric potential inside the box and, along with $\zeta$ and $\mu$ from (\ref{eqn:ass:zeta,mu}), can be substituted in (\ref{eqn:mu}) and (\ref{eqn:zeta}) to obtain the chemical potential derivative $\partial_t\mu$ and the electrodeposition rate $\partial_t\zeta$ {within the transition zone $\zeta\in(\zeta_0,\zeta_1)$}. We use $\phi$, $\partial_t\mu$ and $\partial_t\zeta$ to define descriptors.

\begin{figure}
     \centering
     \begin{subfigure}[b]{0.4\textwidth}
        \centering
        \includegraphics[width=1\linewidth]{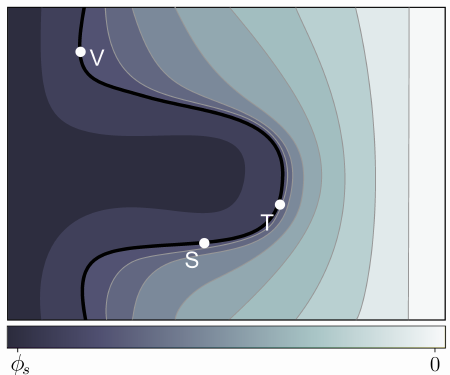}
	    \caption{Curved electric potential level sets promote dendrite growth.}
         \label{fig:dendrite1}
     \end{subfigure}
    \hfill
    \begin{subfigure}[b]{0.4\textwidth}
         \centering
        \includegraphics[width=1\linewidth]{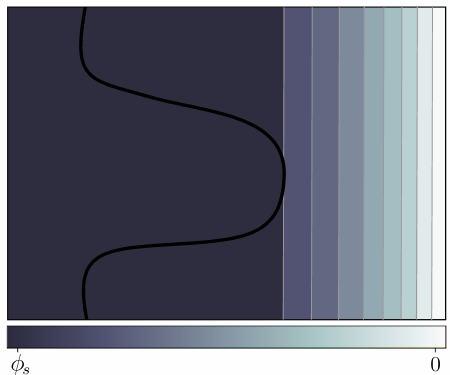}
	    \caption{Linear electric potential level sets inhibit dendrite growth.}
       \label{fig:dendrite2}
    \end{subfigure}
        \caption{The electric potential level sets in a charging half cell, where the anode is on the left and the electrolyte is on the right {($\phi_s<0$). The black curve shows the electrode/electrolyte interface, which is \emph{sharp} unlike in Figure~\ref{fig:box}.}}
        \label{fig:dendrite}
\end{figure}

\textcolor{newred}{First, we define a descriptor for the electric potential.} The electrodeposition process curves the electric potential level sets forward. This phenomenon, observed in numerous studies on different phase-field models (see, \emph{e.g.}, \cite{zinc3,chen2015modulation,ren2020inhibit}), has been acknowledged as a contributing factor to further dendrite growth~\cite{hong2018phase,zinc1,18,40}. Figure~\ref{fig:dendrite} demonstrates this effect by showing a half cell with a grown dendrite (black curve) being charged in a curved electric field (Figure~\ref{fig:dendrite1}) versus a straight electric field (Figure~\ref{fig:dendrite2}). As the electric potential level sets are orthogonal to the electric field, the cations $M^{n+}$ are forced toward the dendrite in the curved field, resulting in its further growth (Figure~\ref{fig:dendrite1}). In contrast, in an ideal case where the electric potential level sets are straight, the dendrite cannot grow larger, because the electric force direction is unchanged throughout the half cell, and thereby, the ions $M^{n+}$ tend to move straight toward the electrode (see Figure~\ref{fig:dendrite2}). Therefore, a less curved electric field can inhibit dendrite growth. Whether the electric field in Figure~\ref{fig:dendrite} is curved or not can be detected by observing the electric potential inside a small box {within} the interface, similar to the one shown in Figure~\ref{fig:box}, at either of the points $S$ (side), $T$ (tip), or $V$ (valley). In the curved field (Figure~\ref{fig:dendrite1}), the electric potential changes more rapidly on the liquid side compared to the solid side. However, in the straight field (Figure~\ref{fig:dendrite2}), the electric potential changes uniformly across the interface in both boxes. Let $\phi(\zeta)$ be the solution to (\ref{eqn:phi_ODE}) and
\begin{equation}\label{eqn:d^phi}
    d^{\phi}:=\phi\left(\frac{\zeta_0+\zeta_1}{2}\right)
\end{equation}
denote the {electric} potential at the interface center. Since
\begin{align*}
d^{\phi}&=\phi_0+\int_{\zeta_0}^{(\zeta_0+\zeta_1)/2}\phi'(\zeta)d\zeta \\
&=\phi_1-\int_{(\zeta_0+\zeta_1)/2}^{\zeta_1}\phi'(\zeta)d\zeta,
\end{align*}
assuming a decreasing electric potential across the interface and the same boundary conditions (\ref{eqn:2pbc}), the cell with a smaller $d^{\phi}$ has a larger total absolute change in $\phi$ on the liquid side of the interface ($\zeta\in[\zeta_0,(\zeta_0+\zeta_1)/2]$) and a smaller one on the solid side ($\zeta\in[(\zeta_0+\zeta_1)/2,\zeta_1]$). Therefore, a more severe dendrite growth is expected when $d^{\phi}$ is small. Straightness of the electric field is not the only factor that suppresses dendrite formation. A more common way to inhibit dendrites is through the chemical potential buildup near the electrode surface, which is introduced next. 

\begin{figure}
     \centering
     \begin{subfigure}[b]{0.4\textwidth}
        \centering
        \includegraphics[width=1\linewidth]{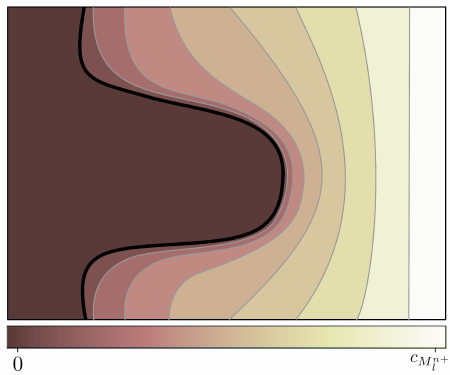}
	    \caption{A decreasing concentration toward the electrode promotes dendrite growth. In this case, the chemical potential is \emph{lower} close to the interface than the bulk electrolyte.}
         \label{fig:dendrite_mu1}
     \end{subfigure}
    \hfill
    \begin{subfigure}[b]{0.4\textwidth}
         \centering
        \includegraphics[width=1\linewidth]{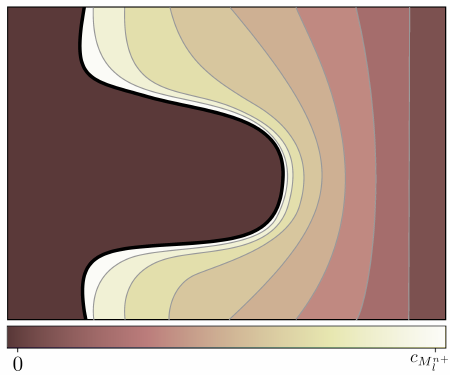}
	    \caption{An increasing concentration toward the electrode inhibits dendrite growth. In this case, the chemical potential is \emph{higher} close to the interface than the bulk electrolyte.}
       \label{fig:dendrite_mu2}
    \end{subfigure}
        \caption{Ion concentration (molar ratio) level sets in a charging half cell, where the anode is on the left and the electrolyte is on the right {($0<c_{M^{n+}_l}$). The black curve shows the electrode/electrolyte interface, which is \emph{sharp} unlike in Figure~\ref{fig:box}.}}
        \label{fig:dendrite_mu}
\end{figure}

Diffusion of ions through the electrolyte often occurs at a much slower rate than their reduction and adsorption on the electrode surface. This condition promotes dendrite growth by consuming the ions $M^{n+}$ at the first available spots as soon as they reach the electrode surface. Saturating the reaction sites with an accumulation of ions near the electrode surface can help prevent this condition~\cite{mueffect}. Slowing the adsorption rate or speeding up the diffusion process can help maintain an accumulation of ions near the electrode surface. In the ideal case, this accumulation of ions would reverse the concentration gradient inside the cell and inhibit dendrite growth. The reason is that the ions are forced towards regions with a lower concentration, which counteracts the electric force that promotes dendrite growth in Figure~\ref{fig:dendrite1}. Figure~\ref{fig:dendrite_mu} shows what the concentration field looks like when a half cell is charged in these two scenarios. {The former case is shown in Figure~\ref{fig:dendrite_mu1}, where} the local concentration decreases when moving from the bulk electrolyte toward the electrode. In contrast, in the latter case shown in Figure~\ref{fig:dendrite_mu2}, the local concentration near the electrode is higher than that in the bulk electrolyte. Note that there is a monotonic relation between concentration $c_{M^{n+}}=c^l(\mu)$ and chemical potential $\mu$ given a fixed phase order (see~(\ref{eqn:phasefield_2})). Hence, the accumulation of ions in Figure~\ref{fig:dendrite_mu2} manifests as a chemical potential buildup close to the electrode surface. A high chemical potential growth rate within the electrode/electrolyte interface implies accelerated charging in the local model. To examine this, one may solve (\ref{eqn:phi_ODE}) for $\phi$ and substitute the solution in (\ref{eqn:mu}) to obtain the chemical potential time-derivatives inside the box. Then we define the descriptor
\begin{equation}\label{eqn:d^mu}
    d^{\mu}:=\frac{1}{\zeta_1-\zeta_0}\int_{\zeta_0}^{\zeta_1} \partial_t \mu d\zeta,
\end{equation}
which measures the average growth rate of chemical potential within the interface. A larger $d^{\mu}$ indicates a faster plating within the interface and therefore, a faster charging process. 

\textcolor{newred}{Finally, we introduce descriptors to evaluate electrodeposition rates.}
Electrodeposition mainly occurs {within} the interface, though at different rates, depending on the region. Typically, a valley region (point $V$ in Figure~\ref{fig:dendrite1}) grows more slowly than a tip region (point $T$ in Figure~\ref{fig:dendrite1}), leading to dendrite growth on the electrode. A similar deposition rate at the tip and valley regions indicates that dendrites do not grow larger. It is possible to evaluate the electrodeposition rates in different regions by using the local model. As can be seen in Figure~\ref{fig:dendrite}, the electric potential tends to vary less across the interface in the valley regions (\emph{e.g.}, point $V$) compared to the tip regions (\emph{e.g.}, point $T$). This phenomenon which is also observed in \cite{separator_effect,zinc1} makes it possible to distinguish a valley from a tip region by the different boundary conditions~(\ref{eqn:2pbc}). Therefore to obtain the electric potential in these two regions, equation (\ref{eqn:phi_ODE}) is solved for $\phi$ using two different boundary conditions that satisfy
$$
\phi^{\textnormal{tip}}_1<\phi^{\textnormal{val}}_1<
\phi^{\textnormal{val}}_0<\phi^{\textnormal{tip}}_0.
$$
Then the solutions are substituted for $\phi$ in (\ref{eqn:zeta}) to give the electrodeposition rates in the tip and valley regions. Let us denote the deposition rates at the tip and valley regions by $\partial_t \zeta^{\textnormal{tip}}$ and $\partial_t \zeta^{\textnormal{val}}$ respectively and define the descriptors
\begin{align*}
        d^{\zeta}_{\textnormal{val}}:=\frac{1}{\zeta_1-\zeta_0}\int_{\zeta_0}^{\zeta_1} \partial_t \zeta^{\textnormal{val}} d\zeta,\\
        d^{\zeta}_{\textnormal{tip}}:=\frac{1}{\zeta_1-\zeta_0}\int_{\zeta_0}^{\zeta_1} \partial_t \zeta^{\textnormal{tip}} d\zeta.
\end{align*}
Dendrites are expected to grow slower in a cell where the quantity $d^{\zeta}_{\textnormal{val}}-d^{\zeta}_{\textnormal{tip}}$ is large while the cell is expected to charge faster when the average deposition rate in the valley and tip regions, \emph{i.e.}, the quantity $d^{\zeta}_{\textnormal{val}}+d^{\zeta}_{\textnormal{tip}}$ is large.



\section{Discussion}\label{sec:discussion}


We defined four different descriptors associated with dendrite growth and charging speed that are summarized in Table~\ref{tab:drule}. We use these descriptors along with the results of Section~\ref{sec:results} to elaborate on the influence of different parameters on the charging process in this section. We consider a lithium-metal half-cell battery with the parameters reported in \cite{hong2018phase}. To calculate the descriptors in Table~\ref{tab:drule} we choose $\zeta_0=0.3$, $\zeta_1=0.7$, $\beta=10$, $\mu_s=-10$ for the local approximation (\ref{eqn:ass:zeta,mu}) (the chosen values are not definitive). 

\begin{table}[t]
\begin{center}
\begin{tabular}{|l l l |} 
 \hline
 Notation & Associated with & Effect\\ [0.5ex] 
 \hline\hline
 $d^\phi$ & Linearity of the electric field & Dendrite inhibition \\ 
 $d^\mu$  & Chemical potential buildup & Fast charging \\ 
$d^{\zeta}_{\textnormal{val}}-d^{\zeta}_{\textnormal{tip}}$  & Deposition uniformity & Dendrite inhibition \\
$d^{\zeta}_{\textnormal{val}}+d^{\zeta}_{\textnormal{tip}}$  & Average deposition rate &
Fast charging
\\
  \hline
\end{tabular}
\caption{Descriptors suggestive of fast charging and dendrite growth. A larger value of these descriptors implies the effects described in the last column.}
\label{tab:drule}
\end{center}
\end{table}

\begin{figure}
	\begin{center}
    \includegraphics[width=1\linewidth]{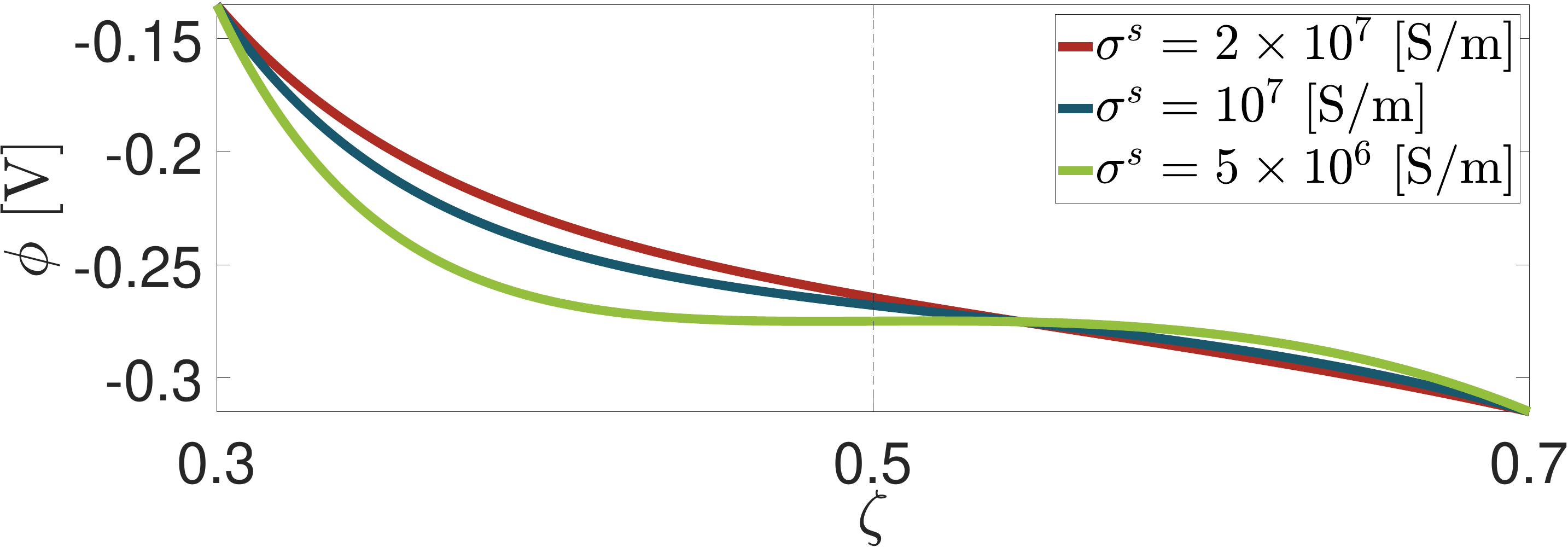} 
	\caption{Electric potential across the phase-transition zone. A larger electrode conductivity $\sigma^s$ yields a straighter electric field across the phase transition zone, and therefore, less dendrite growth.}
	\label{fig:phi}
	\end{center}
\end{figure}

According to Table~\ref{tab:ranges}, a larger electronic conductivity of the electrode $\sigma^s$ can inhibit dendrite growth. This inhibition of dendrites is a consequence of a low electric field curvature. To see this, we note that the conductivities of the electrode and electrolyte are the only parameters that exclusively shape the electric potential via the \textcolor{newred}{conduction} equation (\ref{eqn:phi}). In particular, to reveal the effect of electrode conductivity on the electric field curvature we compute the descriptor $d^{\phi}$ in (\ref{eqn:d^phi}) for three different values of $\sigma^s$. It is observed that a larger conductivity of the electrode results in a larger $d^{\phi}$, and thereby, a more straight electric field. Figure~\ref{fig:phi} shows this effect by plotting the electric field across the interface $\zeta\in[\zeta_0,\zeta_1]$. These results show no trade-off between dendrite inhibition and most other desirable performance objectives in batteries when choosing electrode conductivity $\sigma^s$. There are several ways to change conductivities in a battery cell. One may change the electrode conductivity $\sigma^s$, by \emph{e.g.}, tuning the metal composition (alloying) and the operating temperature. 


In contrast with electrode conductivity, a smaller ionic conductivity of the electrolyte $\sigma^l$ can inhibit dendrite growth, according to Table~\ref{tab:ranges}. This kind of dendrite inhibition comes, however, with a trade-off, as a high ionic conductivity of the electrolyte is desirable for the cell overall performance~\cite[\S1.5.4]{berg2015batteries}. {The suppressive effect of lower ionic conductivity of the electrolyte is supported by findings from Rehnlund et al. reporting that reducing lithium salt concentration decreases ionic conductivity and mitigates dendrite formation by favoring two-dimensional lithium deposition under diffusion-controlled conditions~\cite{Conductivity_application}.} The results in Table~\ref{tab:ranges} also state that the half cell is charged faster when the ionic conductivity of electrolyte is high. This is of course expected, as the half-cell resistance is dominated by the electrolyte resistance and when this resistance is low, the charging current is high under a constant voltage by the Ohm's rule. It is possible to change the electrolyte conductivity $\sigma^l$, by \emph{e.g.}, changing the viscosity, the used solvent, the salt concentration, or the temperature~\cite[\S 1.5.4]{berg2015batteries}. 


\begin{figure}
	\begin{center}
    \includegraphics[width=1\linewidth]{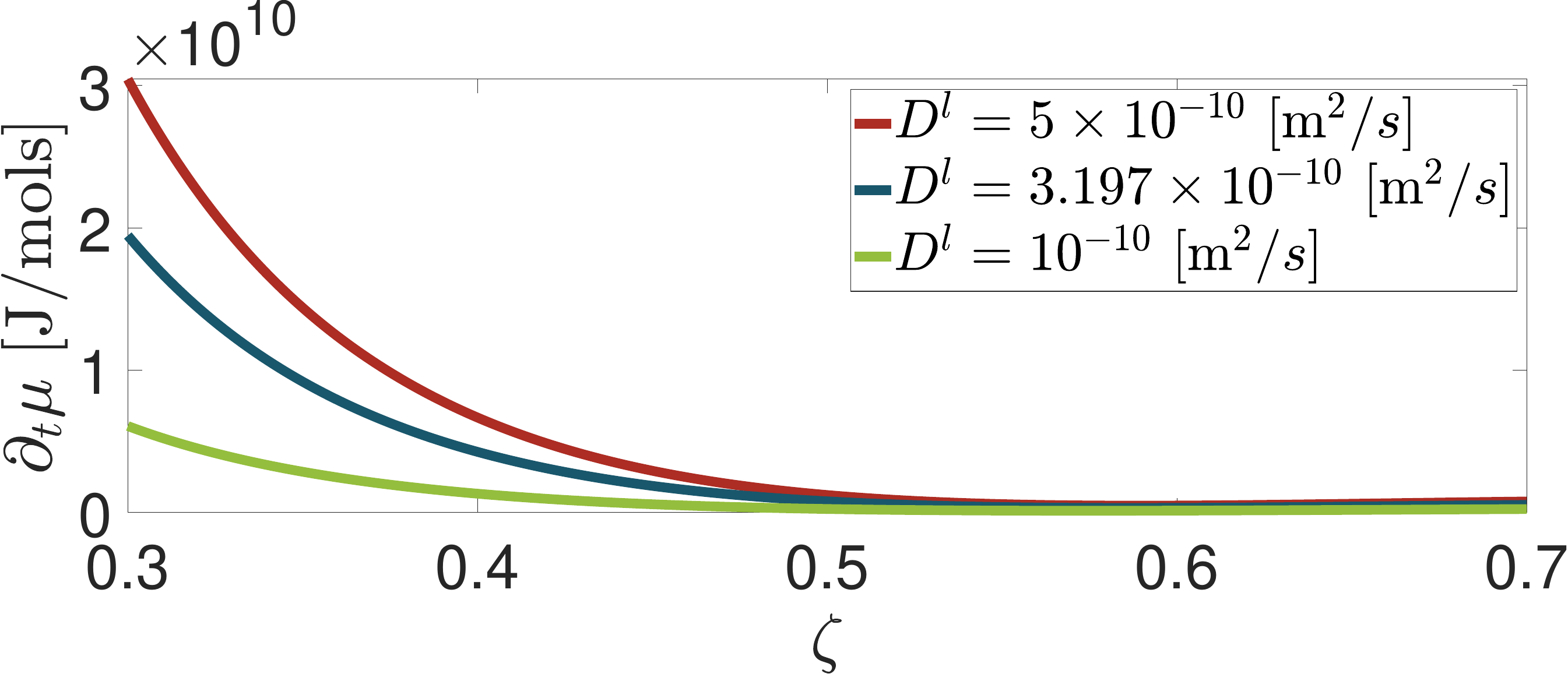}  
	\caption{Chemical potential growth rate across the phase-transition zone. A larger diffusion coefficient $D^l$ yields a higher chemical potential growth rate on average.}
	\label{fig:mu}
	\end{center}
\end{figure}

The only parameters that exclusively influence the chemical potential via the diffusion equation (\ref{eqn:mu}) are the diffusion coefficient $D^l$ and the site density of electrolyte $C_m^l$. To study the effects of the diffusion coefficient on chemical potential, we calculate $d^{\mu}$ for different values of $D^l$ by solving the equation (\ref{eqn:phi_ODE}) and substituting its solution in (\ref{eqn:mu}) to obtain the chemical potential derivatives as shown in Figure~\ref{fig:mu}. It is observed that a larger $D^l$ results in a larger $d^{\mu}$ and therefore, in a chemical potential buildup near the electrode surface. This results in faster charging confirming the results of Table~\ref{tab:ranges}. Table~\ref{tab:ranges} also states that a smaller diffusion coefficient results in less dendrite growth, which is contradictory to most available guidelines that suggest inhibiting dendrites by reaction-limited regimes {\cite{mueffect,zinc2}}. However, a short-term dendrite inhibition as in Table~\ref{tab:ranges} is a consequence of slower ion transport, rather than an accumulation of ions explained in~{\cite{How_dendrites_form}}. 
In contrast, dendrite inhibition in {\cite{mueffect,zinc2}} is the result of an ion accumulation near the interface, which extends the Sand time of the system (see Figure~\ref{fig:dendrite_mu2}). However, an opposite trend appears to hold in the short term, according to Table~\ref{tab:ranges}. Similar to electrolyte conductivity, the diffusion coefficient $D^l$ can also be changed in various ways, including changing the viscosity of the electrolyte.

\textcolor{newred}{
As Table~\ref{tab:ranges} suggests, a lower interfacial energy helps suppress dendrites. This is expected because when the tension is lower, it can be released more easily, which can lead to a smoother surface. In contrast, a high interfacial thickness is beneficial for dendrites suppression, which is in agreement with the results of ~\cite{thickness}. Interestingly, a higher interfacial thickness also promotes the charging speed, according to Table~\ref{tab:ranges}. We will return to this point when discussing interfacial mobility in the next section.
}

According to Table~\ref{tab:ranges}, a lower electrochemical reaction kinetic constant $L_\eta$ can inhibit dendrite growth. The reaction constant $L_\eta$ has a direct relation with the exchange current density $i_0$~\cite{Cogswell} and it is well-documented that a lower current density $i_0$ also inhibits dendrites~\cite{exchange_current_density}, which confirms our results. A lower exchange current density can be achieved via anti-catalysis, \emph{i.e.}, by tailoring specifically the electrode surface composition, \emph{e.g.}, with adsorption of other metal cations. Table~\ref{tab:ranges} also shows that a lower reaction constant $L_\eta$ slows down charging, which is again expected, due to the slower electrochemical reactions. In addition, a direct relation is observed between the descriptor $d^{\mu}$ and $L_\eta$, which confirms this result.

\begin{figure}
     \centering
     \begin{subfigure}[b]{0.49\textwidth}
        \centering
            \includegraphics[width=1\linewidth]{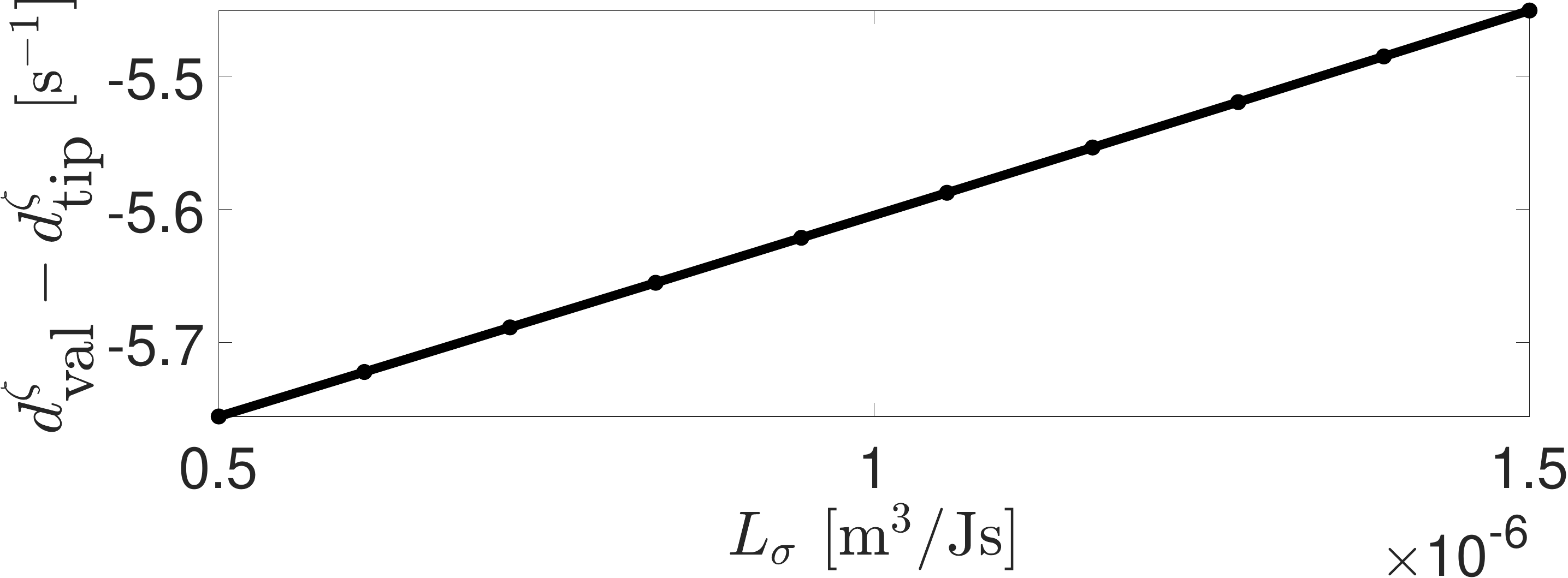}  
	       \caption{Deposition uniformity as a function of interfacial mobility. High interfacial mobility yields a more uniform deposition rate along the electrode surface and hence, less dendrite growth.}
	       \label{fig:dzeta_Lsigma}
     \end{subfigure}
     \hfill
     \begin{subfigure}[b]{0.49\textwidth}
         \centering
                \includegraphics[width=1\linewidth]{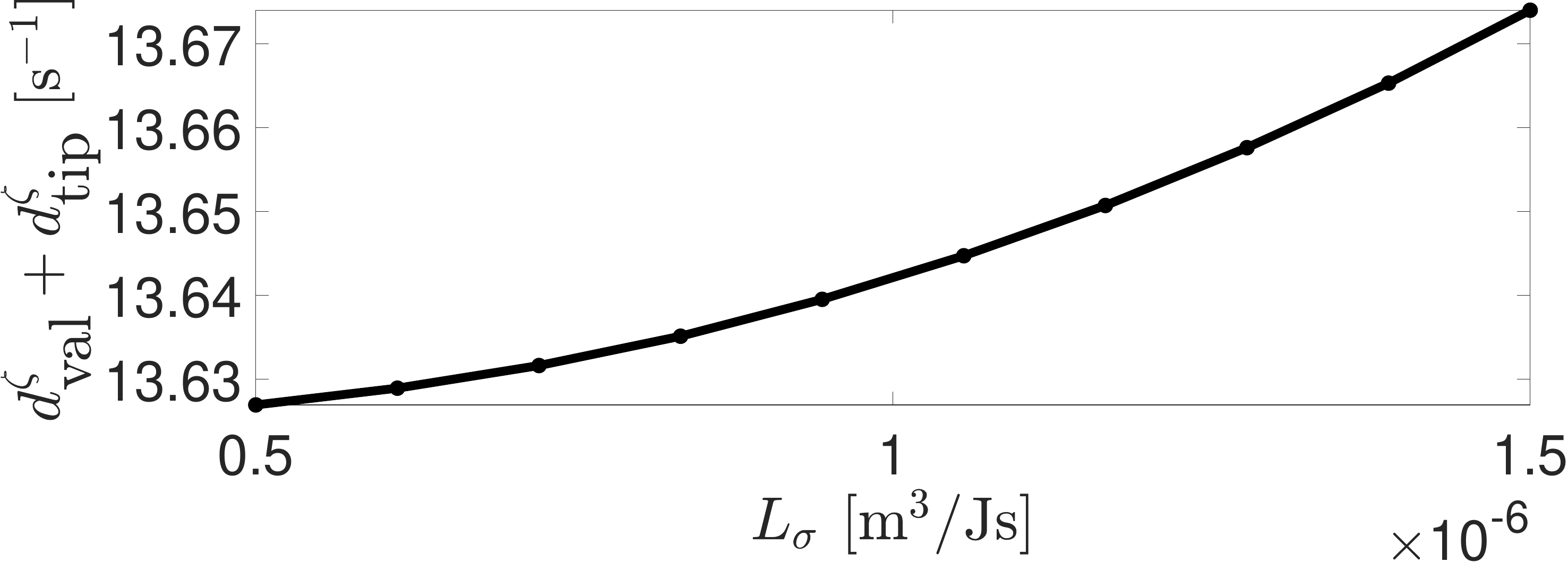} 
	           \caption{Average deposition rate as a function of interfacial mobility. High interfacial mobility yields a faster deposition on average and hence, a higher charging speed.}
	\label{fig:avezeta_Lsigma}
     \end{subfigure}
             \begin{subfigure}[b]{0.49\textwidth}
         \centering
    \includegraphics[width=1\linewidth]{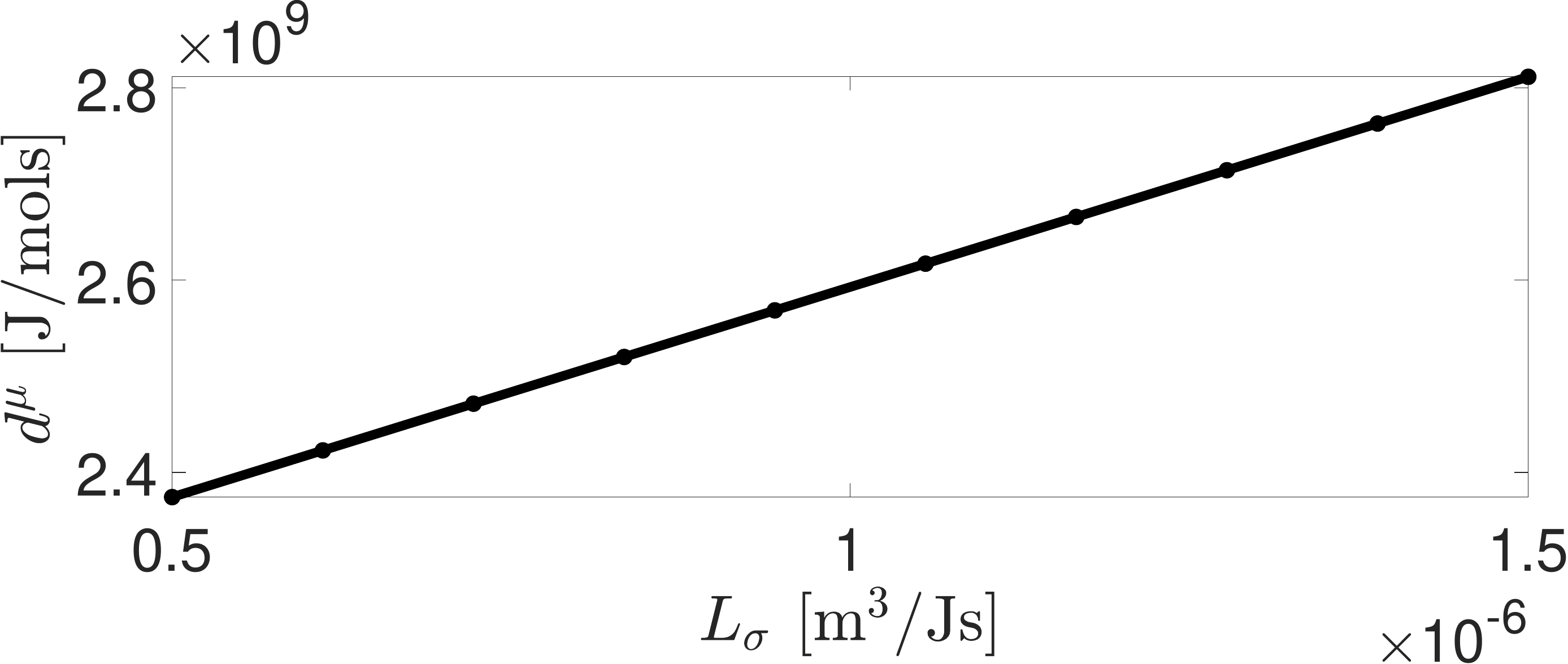} 
	\caption{Chemical potential buildup as a function of interfacial mobility. High interfacial mobility accumulates ions within the interface and builds up the chemical potential there faster, leading to a higher charging speed.}
	\label{fig:dmu_Lsigma}
    \end{subfigure}
    \hfill
    \begin{subfigure}[b]{0.49\textwidth}
         \centering
    \includegraphics[width=1\linewidth]{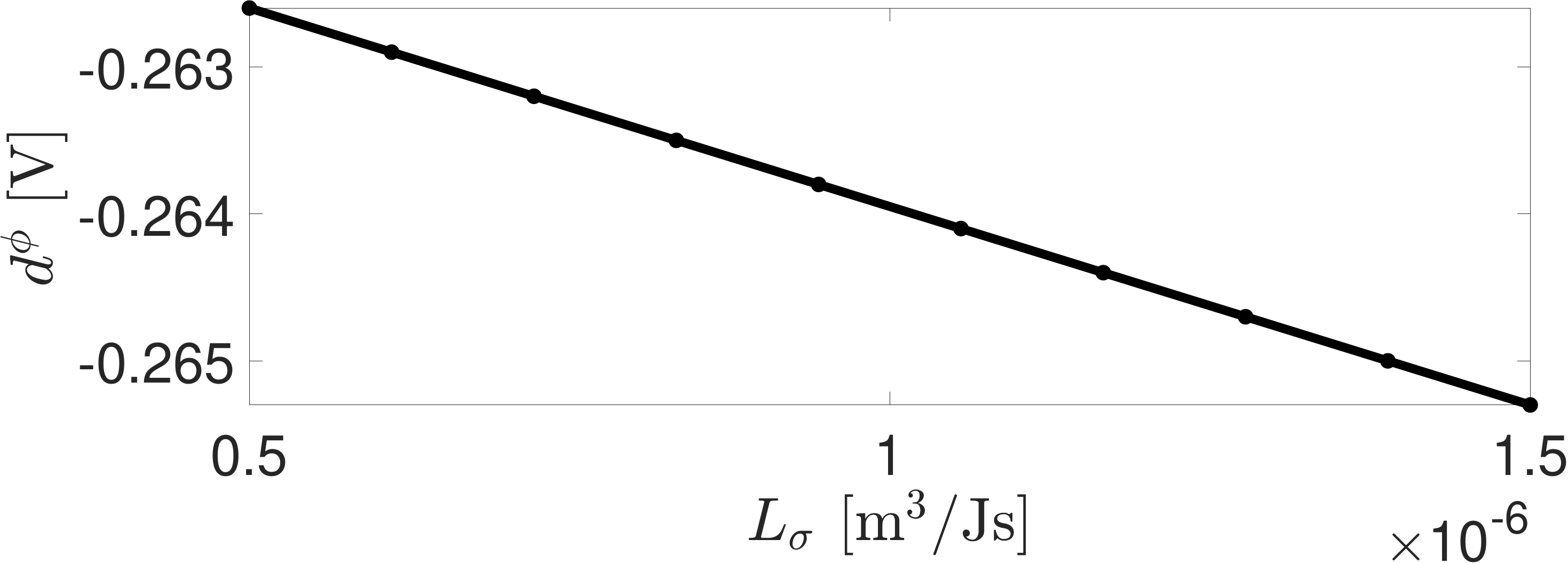} 
	\caption{Linearity of the electric field as a function of interfacial mobility. High interfacial mobility intensifies the electric field curvature, indicating a different mechanism of dendrite inhibition than high electrode conductivity.}
	\label{fig:dphi_Lsigma}
    \end{subfigure}
    \caption{The relation between interfacial mobility $L_\sigma$ and different descriptors defined in Table~\ref{tab:drule}.}
\end{figure}

A high interfacial mobility can inhibit dendrite growth and accelerate charging speed, according to Table~\ref{tab:ranges}. We leverage the descriptors $d^{\zeta}_{\textnormal{val}}$ and $d^{\zeta}_{\textnormal{tip}}$ to study this effect. To compute these descriptors, we solve the differential equation (\ref{eqn:phi_ODE}) separately with the following boundary conditions:
\begin{align*}
\phi(\zeta_0)&=-0.1,\; \phi(\zeta_1)=-0.6,& \quad\textnormal{tip region} \\
\phi(\zeta_0)&=-0.3,\; \phi(\zeta_1)=-0.5,&\quad\textnormal{valley region}
\end{align*}
using different values of $L_\sigma$. The solution $\phi$ is then substituted in (\ref{eqn:zeta}) to obtain the electrodeposition rates in the tip and valley regions. Figures~\ref{fig:dzeta_Lsigma} and \ref{fig:avezeta_Lsigma} show the electrodeposition rate uniformity $d^{\zeta}_{\textnormal{val}}-d^{\zeta}_{\textnormal{tip}}$ and the average electrodeposition rate $d^{\zeta}_{\textnormal{val}}+d^{\zeta}_{\textnormal{tip}}$ along the electrode surface, respectively. Both these descriptors are increasing with $L_\sigma$, which confirms our results in Table~\ref{tab:ranges}. This also agrees with the trend seen in the chemical potential growth rate in Figure~\ref{fig:dmu_Lsigma}. However, an opposite trend is observed in $d^{\phi}$ (Figure~\ref{fig:dphi_Lsigma}), which indicates that the dendrite inhibition achieved by increasing interfacial mobility is not caused by straightening the electric field. Rather, a high interfacial mobility facilitates the surface tension release, which results in a smoother electrode surface and therefore in less dendrite growth.

\textcolor{newred}{Interfacial mobility ($L_\sigma$) has been tuned as a numerical parameter to fit the phase-field simulations to real-world experiments in some studies~\cite{interfacemobility13}. This parameter has a direct relationship with the (intrinsic) interfacial mobility ($\overline{L}_\sigma$), which is a physical parameter of the system. Several papers have approximated this relationship under different conditions, including at the thin interface limit in \cite[Eq.(63)]{interfacemobility}. Perhaps the simplest formula relating the two parameters is derived by \cite{interfacemobility13} as follows
\begin{equation}\label{eqn:intrinsic_int_mobility}
    L_{\sigma}=k_M\frac{\gamma}{k} \overline{L}_\sigma,
\end{equation}
where $\gamma$ is the interfacial energy, $k$ is the gradient energy coefficient~\cite[Eq.(10)]{interfacemobility2}, and $k_M$ is a parameter used to fit the simulations with measurements at different temperatures $T$~\cite[Eq.(23b)]{interfacemobility13}. Recall that $k\propto \gamma\delta$~\cite{Cogswell,ely2014phase}. Hence, equation (\ref{eqn:intrinsic_int_mobility}) yields
\begin{equation}\label{eqn:intrinsic_int_mobility2}
    L_{\sigma}\delta\propto k_M\overline{L}_\sigma.    
\end{equation}
Therefore, with a fixed $k_M$, increasing $L_{\sigma}$ and/or $\delta$ corresponds to increasing $\overline{L}_\sigma$ in the physical system. As we saw earlier in Section~\ref{sec:results}, interfacial mobility and interfacial thickness are the only parameters that are \emph{not} subject to the trade-off observed between charging speed and dendrite suppression. This result suggests that a high intrinsic interfacial mobility can help achieve both objectives in a battery cell.}



\textcolor{newred}{Changing (intrinsic) interfacial mobility is observed in the broader electroplating industry.} In copper electroplating, for example, the control of deposition behavior is largely chemistry-driven, with additives playing a critical role. Suppressors, adsorb on the copper surface and selectively reduce the deposition rate in high-current-density regions like tips, effectively promoting uniform deposition by slowing plating in these areas. Accelerators, counteract the suppressor effect by increasing deposition rates in valley regions where suppressors are less adsorbed, ensuring a balanced deposition and reducing surface roughness. Levelers, acting as secondary suppressors, specifically target protrusions such as dendrite tips to minimize surface irregularities and ensure smoother plating. Together, these additives indirectly adjust $L_\sigma$ and $L_\eta$, with suppressors and levelers mimicking the effects of high $L_\sigma$ by reducing unevenness, while accelerators fine-tune deposition rates to prevent over-suppression~\cite{modern_electroplating}. Similar efforts have been made for lithium with different approaches aimed at improving the interfacial mobility of Li-ions on the surface of the electrode~\cite{Cu_Li_1, Cu_Li_2, Cu_Li_3}.

\textcolor{newred}{We conclude this section by summarizing the main results of this paper.} We have introduced a numerically efficient method for optimizing electrodeposition processes with respect to an arbitrary objective function using an arbitrary set of parameters. This approach is based on Bayesian optimization and uses a numerical solution of the phase-field partial differential equations over a fixed time interval. This time interval is flexible and can be chosen to optimize the short- or long-term behavior of electrodeposition processes. \textcolor{newred}{Choosing a small time horizon reduces the computational burden and enables a faster optimization of the parameters inside the battery cell. However, the optimal parameters found in this way are only reliable for longer charging sessions if there is no dramatic change in the system behavior, such as switching from a reaction-limited regime to a diffusion-limited regime. Nevertheless,} using a low-complexity optimization framework, the proposed approach has great potential for controlling various aspects of electrodeposition processes by tuning the right parameters of battery cells. \textcolor{newred}{This method applies to metal-anode batteries, such as lithium-metal batteries, zinc-metal batteries, and so on, but not batteries that involve ion insertion, such as Li-ion batteries. The method can be easily adapted for different types of batteries by adjusting the corresponding parameters so they reflect the real system, such as the valence $n$ of ions, the electrolyte properties, and so on. We used this framework to obtain the minimum dendrite growth and the maximum charging speed in a lithium-metal battery by tuning the following parameters: electrode conductivity, electrolyte conductivity, site density in electrolyte, diffusion coefficient of electrolyte, interfacial mobility, the electrochemical reaction kinetic coefficient, interface tension, and interfacial thickness. Optimizing other types of batteries under different operating conditions can lead to different conclusions from those drawn in this study.}

The results indicate that dendrite suppression and fast charging are generally conflicting objectives when charging under constant voltage, with the curious exception of one parameter: intrinsic interfacial mobility. Our results show that increasing this parameter can inhibit dendrite growth effectively, without compromising charging speed. This is realized by an easier release of surface tension and an easier adsorption of ions, which reduces the surface roughness and accelerates electrodeposition. \textcolor{newred}{Another interesting observation in our results is that certain information provided in short simulations on dendrites is still reliable for surprisingly long simulation times. Observing tiny differences in how fast or severe dendrites grow in the beginning provides useful information about their size later on. Although the shape of the dendrite is almost impossible to predict, the relation between different parameters and functions (\ref{eqn:ro})--(\ref{eqn:SOC}) did not change in the short and long simulations in most of our experiments. We deduce that, while short-term simulations can not predict everything about the evolution of dendrites (such as dramatic changes in the system behavior caused by low or high diffusion coefficients), they provide cheap but valuable information about how fast they grow. This information can accelerate the search for high-performance batteries.}

To also gain more insight into how different parameters influence the electrodeposition process, we introduced a few simple descriptors by localizing the phase-field model across the electrode/electrolyte interface in an infinitesimal time interval. The reduced model and its descriptors reveal the close relationship between electric and chemical potential fields and the chemical parameters of a battery cell. \textcolor{newred}{The proposed descriptors are easy to compute without simulating the full phase-field model. As such, they make} an initial step toward a physically-inspired computationally-tractable model based on phase-field equations for real-time feedback control systems. This shall be a future direction for further research.

\section{Methods}\label{Sec:Methods}
\subsection{Decision variables}
\textcolor{newred}{The phase-field model used in this paper describes a half-cell with the following dimensions: It spans from the anode location $x_0$ on the left to the electrolyte end $x_f$ on the right, where $x_0<x_f$. The width of the system from the bottom $y_0$ to the top $y_f$ is given by $y_f-y_0$.}  

\textcolor{newred}{This model assumes the following set of constant parameters: interfacial mobility $L_\sigma$, gradient energy coefficient $k$, reaction constant $L_\eta$, the valence of charge carriers $n$, site density of electrolyte $C^l_m$, site density of electrode $C^s_m$, energy barrier height $W$, temperature $T$, the chemical potential differences in the electrolyte (electrode) at initial equilibrium state $\epsilon^l$ ($\epsilon^s$), the standard equilibrium half-cell potential $E^{\theta}$, charge transfer coefficient $\alpha$, diffusion coefficient of the ions $M^{n+}$ in the electrolyte $D^l$, initial molar ratio of ions $c_0$, and finally, the conductivities of electrolyte and electrode $\sigma^l$ and $\sigma^s$.}

\textcolor{newred}{The following parameters are implicit in the model: interfacial energy $\gamma$ and interfacial thickness $\delta$, which control the barrier height $W=12\gamma/\delta$ and the gradient energy coefficient $3\gamma\delta/2$. The following parameters are implicit in the initial and boundary conditions: the applied voltage $\phi_s<0$, which is measured across the half-cell, from the anode (solid phase) to the probe electrode (with zero electric potential), the initial chemical potential in the solid phase $\mu_s<0$, and the initial surface of the electrode given by the functional $\zeta_0:[x_0,x_f]\times [y_0,y_f]\to [0,1]$.}

\textcolor{newred}{Several of the above parameters can be included as decision variables $\theta$ in the optimization problem (\ref{eqn:opt}). The fundamental constants of the model are the Faraday constant $F$ and the gas constant $R$.}

\subsection{Initial and boundary conditions}
\textcolor{newred}{Next, we describe the conditions used to solve the phase-field equations. For boundary conditions, we use the Dirichlet conditions
\begin{align*}
&\zeta(x_0,y,t)=1,\; \zeta(x_f,y,t)=0,\\
&\mu(x_f,y,t)=0,\\
&\phi(x_0,y,t)=\phi_s,\; \phi(x_f,y,t)=0
\end{align*}
and the following conditions of Neumann type
\begin{align*}
&\partial_y\zeta(x,y_0,t)=\partial_y\mu(x,y_0,t)=\partial_y\phi(x,y_0,t)=0,\\
&\partial_y\zeta(x,y_f,t)=\partial_y\mu(x,y_f,t)=\partial_y\phi(x,y_f,t)=0,
\end{align*}
where $\phi_s=-0.45$ is set for all our experiments. The initial conditions, on the other hand, varied among the experiments. We adopted the following initial condition from~\cite{RebeccaCodes} for the experiments used to generate Figures~\ref{fig:posterior} and \ref{fig:lamplot}:
\begin{align*}
\zeta(x,y,0)&=\frac{1}{2}\biggl(1-\tanh\Bigl(2x-40
\\
&+0.35\exp\bigl(-0.1(y-50)^2\bigr)\Bigr)\biggr),\\
\mu(x,y,0)&=\mu_s H(20-x),
\end{align*}
where $\mu_s=-10$ and $H(.)$ is the Heaviside step function. For the rest of the experiments in Section~\ref{sec:results}, we use the more intricate initial condition shown in Figure~\ref{fig:slow_full}.}

\subsection{Hardware}
\textcolor{newred}{Solving the Bayesian optimization problems and simulating the complete phase-field models were carried out on the Tetralith supercomputer at the National Supercomputer Centre (NSC) with 32 cores. The computing time varied for different experiments. For instance, the second experiment of Section~\ref{sec:results} took 24 hours 13 minutes and 17 seconds to complete, and the simulation performed to generate Figure~\ref{fig:Lsigma} took 5 days, 4 hours, 47 minutes, and 1 second to complete on this machine.}

\subsection{Software}
\textcolor{newred}{The two-point boundary value problem (18) was solved by the four-stage Lobatto IIIa collocation formula using the Matlab built-in function bvp5c. The system of partial differential equations (1)-(3) was numerically solved by the finite element method using the Fenics software. We built upon the work~\cite{RebeccaCodes} for this purpose and chose Newton's method to solve the variational problems. To solve the Bayesian optimization problem (12), we used the Ax-BoTorch platform in all experiments~\cite{balandat2020botorch}. The prior was initialized by at least $5$ quasi-randomly generated trials. For the kernel function, we used Matern 5/2 function as follows
$$
K(\theta,\theta')=\frac{2^{1-\nu}(2\nu)^{\nu/2}}{\Gamma(\nu)}\vert\theta-\theta'\vert^\nu K_\nu\bigl(\sqrt{2\nu }\vert\theta-\theta'\vert\bigr),
$$
where $K_\nu$ is the modified Bessel function of the second
kind and $\nu=5/2$. We used sequential Bayesian optimization for all the experiments, with only one sample used to evaluate the posterior at each step. The number of trials used for each of the first experiments in Section~\ref{sec:results} was $20$, for the second experiment in Section~\ref{sec:results} was $100$, and for each experiment performed to generate column $\lambda=0$ in Table~\ref{tab:ranges} was $20$. The number of trials used to obtain the optimal reaction constants in Figure~\ref{fig:lamplot} was $10$ for each value of $\lambda$. For the acquisition function, we considered expected improvement, defined as follows
\begin{align*}
    A(\theta\,\vert\, \mathcal{D}_0,\mathcal{D}_1,\dots,\mathcal{D}_i)&=\mathbb{E}\bigl(\max(0,C(t_f;\theta)-C^\star_i)\bigr)\\
&:=\textnormal{EI}_i(\theta),
\end{align*}
where $C^\star_i=\max_{j\leq i}C(t_f;\theta_j)$ is the best objective function value observed so far. In practice, the logarithm of the noisy version of this acquisition function is optimized instead for its numerical benefits using gradient-based algorithms~\cite{ament2023unexpected}. 
}

\section*{Data availability}
The detailed parameters values used for simulations and the datasets generated during the current study are available in the repository~\cite{github}.

\section*{Code availability}
The underlying codes for this study are available in the repository~\cite{github}.




\section*{acknowledgement}
This work was funded by the National Strategic e-Science Program eSSENCE, the Göran Gustafsson Foundation, and the Knut and Alice Wallenberg Foundation through the Wallenberg Initiative Material Science for Sustainability (WISE) and the Wallenberg AI, Autonomous Systems and Software Program (WASP). Stiftelsen för Strategisk Forskning (SSF; FFL18-0269) and Knut and Alice Wallenberg (KAW) Foundation (grant 2017.0204) are acknowledged for project funding and STandUp for Energy for base funding. Also, The Governmental Intiative for Excellence and Education for the Electrification of the Transport Sector using Battery Technologies (COMPEL) is acknowledged for funding. The computations were enabled by resources provided by the National Academic Infrastructure for Supercomputing in Sweden (NAISS), partially funded by the Swedish Research Council through grant agreement no. 2022-06725.  None of the funders played any role in the study design, data collection, analysis and interpretation of data, or the writing of this manuscript.

\section*{Author contributions}
HT: Writing - Original Draft, Software, Visualization, Methodology, Investigation, Formal analysis.
VV: Writing - Review \& Editing, Validation. 
EB: Writing - Review \& Editing, Project administration, Supervision, Funding acquisition. 
PB: Writing - Review \& Editing, Project administration, Resources. 
JS: Writing - Review \& Editing, Project administration, Supervision, Funding acquisition.

\section*{Competing interests}
All authors declare no financial or non-financial competing interests.

\bibliography{main.bib}
\bibliographystyle{naturemag}

\end{document}